\RequirePackage{rotating}

\documentclass[11pt,a4paper,fleqn]{article}
\usepackage[utf8]{inputenc}
\usepackage{authblk}
\usepackage{geometry}
\usepackage{xcolor}
\usepackage[round]{natbib}
\usepackage{graphicx}
\usepackage{pdflscape}  
\usepackage{appendix}
\usepackage{threeparttable}
\usepackage{booktabs}
\usepackage{subcaption}
\usepackage{longtable}
\usepackage{amssymb}
\usepackage{amsbsy}
\usepackage{bm}
\usepackage{amsmath}
\usepackage{amsthm}
\usepackage{graphicx}
\usepackage{mathtools}
\usepackage{setspace}
\usepackage{color, colortbl}
\definecolor{ashgrey}{rgb}{0.7, 0.75, 0.71}
\definecolor{columbiablue}{rgb}{0.61, 0.87, 1.0}
\definecolor{coral}{rgb}{1.0, 0.5, 0.31}

\definecolor{colBVAR}{HTML}{bababa}
\definecolor{colBART}{HTML}{d7191c}
\definecolor{colmixBART}{HTML}{fdae61}
\definecolor{colerrorBART}{HTML}{abd9e9}
\definecolor{colfullBART}{HTML}{2c7bb6}

\definecolor{colcons}{HTML}{e31a1c}
\definecolor{colSV}{HTML}{a6cee3}
\definecolor{colhBART}{HTML}{1f78b4}

\usepackage{enumitem}
\newlist{steps}{enumerate}{1}
\setlist[steps,1]{label = Step \arabic*:}

\usepackage{adjustbox}
\usepackage{dcolumn}
\newcolumntype{d}[1]{D..{#1}} 

\usepackage{xcolor,colortbl}

\definecolor{Gray}{gray}{0.85}
\definecolor{LightCyan}{rgb}{0.88,1,1}

\newcolumntype{a}{>{\columncolor{Gray}}c}
\newcolumntype{b}{>{\columncolor{white}}c}

\usepackage{caption}
\usepackage{subcaption}
\definecolor{nblue}{HTML}{000660}
\usepackage[colorlinks=true,urlcolor=nblue,linkcolor=nblue,citecolor=nblue]{hyperref}
\newcommand*{\myeqref}[2][Eq.~]{%
  \hyperref[{#2}]{#1(\ref*{#2})}%
}
\def\equationautorefname#1#2\null{%
  Eq.#1(#2\null)%
}

\begin{document}
\title{\textbf{Subspace Shrinkage in Conjugate Bayesian Vector Autoregressions}\thanks{
We would like to thank Michael Pfarrhofer and Niko Hauzenberger for constructive comments and suggestions. Huber gratefully acknowledges financial support from the Austrian Science Fund (FWF, grant no. ZK 35) and the Jubiläumsfonds of the Oesterreichische Nationalbank (OeNB, grant no. 18304). }}

\author[a]{Florian \textsc{Huber}}
\author[b]{Gary \textsc{Koop}}
\affil[a]{\textit{University of Salzburg}}
\affil[b]{\textit{University of Strathclyde}}
\date{\today}

\maketitle\thispagestyle{empty}\normalsize\vspace*{-2em}\small\linespread{1.5}
\begin{center}
\begin{minipage}{0.8\textwidth}
\begin{center}
    \small \textbf{Abstract}\\
\end{center}\vspace{-0.3cm}
 \noindent Macroeconomists using large datasets often face the choice of working with either a large Vector Autoregression (VAR) or a factor model. In this paper, we develop methods for combining the two using a subspace shrinkage prior. Subspace priors shrink towards a class of functions rather than  directly forcing the parameters of a model towards some pre-specified location. We develop a conjugate VAR prior which shrinks towards the subspace which is defined by a factor model.  Our approach allows for estimating the strength of the shrinkage as well as the number of factors. After establishing the theoretical properties of our proposed prior, we carry out simulations and apply it to US macroeconomic data. Using simulations we show that our framework successfully detects the number of factors. In a forecasting exercise involving a large macroeconomic data set we find that combining VARs with factor models using our prior can lead to forecast improvements.    
\\\\ 
\textbf{JEL Codes}: C11, C32, C53

\textbf{Keywords}: Subspace shrinkage, reduced-rank regression, Bayesian VAR
\end{minipage}
\end{center}


\normalsize\newpage
\section{Introduction}
Macroeconomists are increasingly working with multivariate time series models involving large numbers of variables. Traditionally dynamic factor models (DFMs) or Factor Augmented Vector Autoregressions (FAVARs) have been used (see e.g. \cite{geweke1977}, \cite{stock2002macroeconomic}, \cite{BBE} and many others). These typically involve the use of principal components methods to extract the information in the large number of variables into a small number of factors thus avoiding over-parameterization concerns. Starting with   \cite{BanburaGiannoneReichlin:2010:jae:large:bvar} many researchers have been simply including all the variables in a Vector Autoregression (VAR) and using Bayesian  shrinkage priors to avoid over-fitting (see, among many others, \cite{Koop:2013:jae:medium:bvar}, \cite{GiannoneLenzaPrimiceri2015RESTAT}, \cite{Jarocinski2017}, \cite{CarrieroClarkMarcellino:2019:joe:VAR:triang}, \cite{KoopKorobilis2019}, \cite{korobilis2019adaptive}, \cite{giannone2019priors}, \cite{huber2019adaptive}, \cite{Chan2020} and \cite{hauzenberger2021combining}). 

How should the researcher decide whether to use a factor model or a large Bayesian VAR? This question can be answered through a comparison of their predictive performance in a pseudo out of sample forecasting exercise. Alternatively, marginal likelihoods can be used. But pseudo out of sample forecasting evaluation can be time consuming and marginal likelihoods can be sensitive to the prior used. In this paper, we develop an alternative method for choosing between factor models and large VARs. 

But why is there  a need to choose between them when something in between might lead to better forecast performance? This is another question addressed in this paper. We propose a model which shrinks the VAR coefficients towards the implied coefficients of a factor model leading to a model which combines the two. We do so using a subspace shrinkage prior, see \cite{subspace}. A conventional prior shrinks the posterior of a coefficient towards its prior mean, which is typically zero. In contrast a subspace shrinkage prior is a prior on function spaces that shrinks towards a class of functions.  In the present paper, we choose the class of functions to be a factor model such as the FAVAR or DFM. We stay in the class of conjugate priors (although we will discuss how other VAR priors can be accommodated) and, thus, our methods are simple to implement. They do not require the use of computationally-demanding Markov Chain Monte Carlo (MCMC) methods, implying that these techniques are useful in very high dimensional models. We develop a method for estimating the weight put on the DFM restrictions and the number of factors involved in these restrictions. The result is a model which combines the large VAR with a DFM in an optimal way. Alternatively, output from our model can be used to select between the large VAR and the DFM and, if the latter is selected, determine the number of factors. 

We consider two versions of our subspace VAR prior. First, the subspace prior can be combined with a conventional informative VAR prior such as the popular Minnesota prior (see \cite{Minnesota1}, \cite{Minnesota2}, \cite{kadiyala1997numerical}, \cite{sims1998bayesian} and \cite{BanburaGiannoneReichlin:2010:jae:large:bvar} for a natural conjugate implementation). We demonstrate that results from such a model can be interpreted as a weighted average of the Minnesota prior VAR and the factor model. Second, the subspace prior can be combined with a non-informative VAR prior. The result is a new Bayesian VAR prior. In contrast to conventional priors which shrink towards plausible values for the VAR coefficients, our new prior shrinks towards the factor model.    

Our approach is illustrated using synthetic as well as real data. In simulations, we show that our framework accurately detects the number of factors if the true number of factors is small. This finding is independent of the model size. In larger dimensions, and for a larger number of true factors, our model slightly underestimates the true number of factors. To investigate the merits of our approach we apply it to US macroeconomic data. In a forecasting exercise, the different priors which shrink the VAR towards a factor model improve upon a standard BVAR and the DFM. These improvements are pronounced during the global financial crisis and the Covid-19 pandemic.

The remainder of the paper is structured as follows. Section \ref{sec: econometrics} introduces the econometric framework. After providing a brief overview on conjugate Bayesian VARs and DFMs in Sub-section \ref{sec: conj.vars}, we discuss our subspace shrinkage prior which can be used to force the coefficients of the VAR towards the restrictions implied by a DFM in Sub-sections \ref{sec: sub.vars} and \ref{sec: sub.minn.vars}. Sub-section \ref{sec: factorestimation} discusses how our approach can be used to estimate the number of factors alongside the remaining model parameters and provides a brief overview on posterior simulation. Section \ref{sec: appl} applies our techniques to a big US macroeconomic dataset and illustrates its favorable forecasting properties. Section \ref{sec_extensions} discusses how alternative Bayesian VAR priors and  extensions such as stochastic volatility can be incorporated in our techniques.  The final section summarizes and concludes the paper. 

\section{Subspace Shrinkage in VARs}\label{sec: econometrics}
\subsection{Conjugate Bayesian VARs and Dynamic Factor Models}\label{sec: conj.vars}
Let $\{\bm Y_t\}_{t=1}^T$ denote an $M$-dimensional vector of macroeconomic and financial quantities. The number of time series can be large and, in addition, display substantial co-movements. One popular approach of modeling this panel of time series is to assume  $\bm Y_t$ to follow a VAR($p$) process:
\begin{equation}
\bm Y_t = \bm A_1 \bm Y_{t-1} + \dots + \bm A_p \bm Y_{t-p} + \bm \varepsilon_t, \label{eq: VAR}
\end{equation}
where $\bm A_j~(j=1,\dots,p)$ is an $M \times M$-dimensional coefficient matrix and $\bm \varepsilon_t$ is a zero mean Gaussian shock with variance-covariance matrix $\bm \Sigma$.\footnote{For brevity we exclude deterministic terms. In our empirical work we include an intercept.} Equation (\ref{eq: VAR}) can be written as a multivariate regression model as follows:
\begin{equation*}
\bm Y_t = \bm A' \bm X_t + \bm \varepsilon_t,
\end{equation*}
where $\bm A = (\bm A_1, \dots, \bm A_p)'$ and $\bm X_t = (\bm Y'_{t-1}, \dots, \bm Y'_{t-p})'$ denote $K (=Mp) \times M$ and $K \times 1$ matrices, respectively. Stacking $\bm Y_t, \bm X_t$ and $\bm \varepsilon_t$ allows us to recast the model in full-data form:
\begin{equation}
\bm Y = \bm X \bm A + \bm \varepsilon, \label{eq: VARfull}
\end{equation}
with typical $t^{th}$ row of $\bm Y$ given by $\bm Y'_t$, of $\bm X$ given by $\bm X'_t$ and of $\bm \varepsilon$ a typical row is $\bm \varepsilon'_t$. 

Notice that the number of VAR coefficients in $\bm a = \text{vec}(\bm A)$ is $k = (K M)$, which sharply increases with the number of endogenous variables and/or the number of lags. Since $T$ is  moderate for typical macroeconomic datasets, shrinkage is necessary to obtain well behaved estimates and to rule out implausible regions of the parameter space (e.g. regions which would imply explosive roots of the VAR process). 

Bayesian priors on $\bm a$ are often used to provide such shrinkage. If $M$ is large, natural conjugate priors are popular since they allow for fast computation. This arises because they preserve a convenient Kronecker structure for the posterior covariance matrix of $\bm a$, see \cite{Chan2020}. The conjugate prior on $\bm a$ is specified conditionally on $\bm \Sigma$ and takes a Gaussian form:
\begin{equation}
\bm a|\bm \Sigma \sim \mathcal{N}(\text{vec}(\underline{\bm A}), \bm \Sigma \otimes \underline{\bm V}). \label{eq: natconjprior}
\end{equation}
Here, we let $\underline{\bm A}$ denote a prior mean matrix of dimension $K \times M$ and $\underline{\bm V}$ is a $K \times K$ matrix.  The full conditional posterior distribution of $\bm a$ is also Gaussian with:
\begin{align*}
\bm a &| \bm \Sigma, \bm Y \sim \mathcal{N}(\text{vec} (\overline{\bm A}), \bm \Sigma \otimes \overline{\bm V}),\\
\overline{\bm V} &= (\bm X' \bm X + \underline{\bm V}^{-1})^{-1},\quad \overline{\bm A} =  \overline{\bm V} (\bm X' \bm Y + \underline{\bm V}^{-1} \underline{\bm A}).
\end{align*}
The prior on $\bm \Sigma$ is inverted Wishart with prior degrees of freedom $\underline{\nu}$ and scaling matrix $\underline{\bm S}$ which, when combined with the likelihood, yields a marginal posterior which also follows an inverted Wishart distribution whose posterior moments take a standard form (see, e.g., chapter 21 of \cite{BEM2}).

A conventional Bayesian VAR prior such as the Minnesota prior would make particular choices for $\underline{\bm A}$ and $\underline{\bm V}$. An alternative would be to exploit the fact that the data might feature a factor structure. That is, the information in $\bm X$ might be characterized by a small number of $q~ (q \ll K)$ latent factors. These can be estimated using principal components (PCs) which can be implemented through a singular value decomposition (SVD). The SVD allows to decompose $\bm X = \bm F_q \bm L_q'$  in terms of a $T \times q$ matrix $\bm F_q$, which are the estimated factors, and a $K \times q$ matrix $\bm L_q$, which is a matrix of factor loadings.   If the matrix $\bm X$ is of rank $q$, this equation is exact. In general, if the rank of $\bm X$ exceeds $q$, $\bm F_q \bm L_q'$ approximates $\bm X$. Replacing $\bm X$ with $\bm F_q \bm L_q'$ in \autoref{eq: VARfull} shows that the corresponding matrix of regression coefficients $\bm B = \bm L'_q \bm A$ is of dimension $q \times M$, a substantial reduction in the dimension of the state space.
Using the Moore-Penrose inverse of $\bm L_q$, $\bm L_q^{\dagger}$, allows us to express $\bm A$ in terms of $\bm B$ and the estimated loadings:
\begin{equation*}
\bm A = (\bm L_q^{\dagger}) \bm B.
\end{equation*}
This equation enables us to think about a DFM in terms of an otherwise unrestricted VAR with specific restrictions (which are driven by $\bm L_q$) on the VAR coefficients. In a conventional DFM, these restrictions are always dogmatically imposed. In this paper, our goal is to introduce a shrinkage prior which softly pushes the elements in $\bm A$ towards the implied restrictions of the PC regression model.

\subsection{Shrinking the flat prior VAR towards a factor model}\label{sec: sub.vars}
Shrinking the regression model towards a subspace spanned by, e.g., the principal components can be done in several ways. For instance, \cite{Oman1982} show how shrinkage estimators can be used to force an unrestricted regression model towards a projection on a subspace (such as the one spanned by the PCs) as opposed to the origin. This approach uses the eigenvalues of $\bm X' \bm X$ to shrink coefficients towards the space spanned by the first $q$ eigenvectors. Our approach is similar but relies on a modified variant of the functional Horseshoe prior stipulated in \cite{subspace}. This is achieved by setting $\underline{\bm A}=\bm 0_{K \times M}$ and $\underline{\bm  V}$ as follows:
\begin{equation*}
\underline{\bm V} =  \left(\frac{\omega}{1-\omega} \bm X' (\bm I_T - \bm \Phi_0) \bm X \right)^{-1}.
\end{equation*}
Here, $\omega \in [0, 1]$ is a tightness parameter and the $T \times T$ matrix $\bm \Phi_0 = \bm F_q (\bm F_q' \bm F_q)^{-1} \bm F_q' $ is the projection of $\bm F_q$.  Recall that we obtain $\bm F_q$ from the SVD of $\bm X$.\footnote{Note that if we were to set $\bm \Phi_0 = \bm 0_{T \times T}$ then the prior would reduce to a standard $g$-prior with hyperparameter $\omega / (1-\omega)$.} We let $\omega$ be an unknown parameter and estimate it in a data-based fashion as described below.    The posterior is given in the preceding sub-section with these particular choices of $\underline{\bm A}$ and $\underline{\bm  V}$ inserted.

To see how $\omega$ shrinks the VAR towards the factor model, it is convenient to exploit the fact that if the rank of $\bm X$ is $K$, the matrix $\bm X$ and the matrix $\bm F_K$ (i.e., the first $K$ principal components of $\bm X$) span the same column space $C$. Moreover, notice that $\bm F_K = (\bm F_q, \bm F_{(q+1) : K})$ with $\bm F_{(q+1): K}$ storing the final $K-q$ principal components of $\bm X$. Using these definitions and the result that $C(\bm X) = C(\bm F_K)$, the corresponding projection matrices coincide:
\begin{align}
&\bm X \left (\bm X' \bm X + \frac{\omega}{1-\omega} \bm X' (\bm I_T - \bm \Phi_0) \bm X\right)^{-1} \bm X' \\
&= \bm F_K \left (\bm F_K' \bm F_K + \frac{\omega}{1-\omega}  \bm F'_K (\bm I_T - \bm \Phi_0)  \bm F_K \right)^{-1} \bm F_K'. \label{eq: proj_F}
\end{align}
Notice that conditional on a standard normalization, we have that $\bm F'_K \bm F_K = \bm I_K$ and $\bm F_q' \bm F_{(q+1) : K} = \bm 0$. This allows us to rewrite \autoref{eq: proj_F} as:
\begin{align}
&\bm F_K \left (\bm I_K + \frac{\omega}{1-\omega}   \left[\bm I_K -  \begin{pmatrix} \bm I_q & \bm 0 \\ \bm 0 & \bm 0\end{pmatrix}\right]  \right)^{-1} \bm F_K'\\
&= (\bm F_q,~ \bm F_{q+1 : K}) \begin{pmatrix} \bm I_q &\bm 0 \\
\bm 0 &\bm (1-\omega) \bm I_{K-q}  
\end{pmatrix}
\begin{pmatrix}
\bm F'_q \\ \bm F'_{q+1 : K}
\end{pmatrix}\\
&= \bm \Phi_0 + (1-\omega) \bm \Phi_1, \label{eq: convex_0}
\end{align}
with $\bm \Phi_1 = \bm F_{q+1 : K} (\bm F'_{q+1 : K} \bm F_{q+1 : K})^{-1} \bm F'_{q+1 : K}$. It is straightforward to show that
\begin{equation*}
\bm \Phi = \bm \Phi_0 + \bm \Phi_1,
\end{equation*}
where $\bm \Phi = \bm X (\bm X' \bm X)^{-1} \bm X'$ is the projection matrix of $\bm X$. Thus, we can substitute $\bm \Phi_1 = \bm \Phi - \bm \Phi_0$ in \autoref{eq: convex_0} and multiply from the right with $\bm Y$ to arrive at:
\begin{equation}
\mathbb{E}(\bm X \bm A | \bm Y, \omega) = \omega \bm \Phi_0 \bm Y + (1-\omega) \bm \Phi \bm Y, \label{eq: convex}
\end{equation}
which shows that the posterior mean of the regression function is a convex combination of the  VAR fit, $\bm \Phi \bm Y$, and the fit of the PCA regression, $\bm \Phi_0 \bm Y$. This result can be used to show that the resulting predictive distribution (or impulse responses) are weighted averages of the ones obtained from estimating an unrestricted VAR  and a PC regression, both estimated using OLS. Larger values of $\omega$ imply estimates which are closer to the ones obtained from estimating a PC regression while values of $\omega$ closer to zero yield estimates closer to those of a non-informative prior Bayesian VAR.

Note that in the preceding material we are not incorporating any conventional Bayesian VAR prior such as the Minnesota prior. The prior defined above is a new one which can be used if the researcher wishes to use a prior which only shrinks towards the factor model. We will use the acronym subVAR-Flat to denote this prior which combines the subspace prior with a flat prior for the VAR coefficients. The fact that $\omega = 0$ yields  a flat prior VAR illustrates an important aspect and potential shortcoming of this prior. Flat prior VARs tend to over-fit unless $M$ is very small and if $K>T$, as commonly occurs with large VARs, the OLS estimator will not be defined. Adding subspace shrinkage will ensure the posterior is proper, but small values of $\omega$ can potentially lead to over-fitting. As we will document in our empirical results, using a non-informative prior for $\omega$ can lead to poor forecast performance in large VARs. Hence, the need for a suitable prior for $\omega$. This will be provided below.  

\subsection{Shrinking the Minnesota prior VAR towards a factor model}\label{sec: sub.minn.vars}

Since our prior is conjugate, we can easily add additional VAR priors to complement our subspace prior. In this sub-section, we show how this can be done for the natural conjugate Minnesota prior as implemented in \cite{BanburaGiannoneReichlin:2010:jae:large:bvar}.

Let $\overline{\bm Y} = (\bm Y', \underline{\bm Y}')'$ and $\overline{\bm X} = (\bm X', \underline{\bm X}')'$ denote dummy-augmented data matrices. The dummies $\underline{\bm Y}$ and $\underline{\bm X}$ can be specified to match features of the different priors in the Minnesota tradition. We assume that these dummies are parameterized by a hyperparameter $\vartheta$, with values of $\vartheta$ close to zero implying strong shrinkage towards the prior mean. In our empirical work we set the dummies as follows:
\begin{equation*}
    \underline{\bm Y} = \begin{pmatrix}
    \text{diag}(\underline{a}_{1} \hat{\sigma}_1, \dots, \underline{a}_{M} \hat{\sigma}_M)/\vartheta \\
    \bm 0_{M (p-1) \times M} \\
    \text{diag}(\hat{\sigma}_1, \dots, \hat{\sigma}_M) \\
    \bm 0_{1 \times M}
    \end{pmatrix}, \quad \underline{\bm X} = \begin{pmatrix}
    \bm J_p \otimes \text{diag}(\hat{\sigma}_1, \dots, \hat{\sigma}_M)/\vartheta & \bm 0_{K \times 1}\\
    \bm 0_{M \times K} & \bm 0_{M \times 1}\\
    \bm 0_{1 \times K} & \kappa
    \end{pmatrix},
\end{equation*}
with $\hat{\sigma_j}~ (j=1, \dots, M)$ denoting the OLS residual standard deviation of an AR($p$) model for $y_{jt}$, the $j^{th}$ variable in $\bm Y_t$, $\underline{a}_{j}$ is the $j^{th}$ diagonal element of $\underline{\bm A}$, and $\bm J_p = \text{diag}(1, \dots, p)$. Notice that this set of dummies includes the prior for the intercept which depends on the hyperparameter $\kappa$. $\kappa$ is set to a very small number (in our empirical work it equals $0.001$), leading to a weakly informative prior for the intercepts.   

We can add these dummies to $\bm Y$ and $\bm X$ and then combine it with our subspace shrinkage prior. The posterior covariance matrix of the VAR coefficients then becomes:
\begin{equation*}
\overline{\bm V} = \left(\overline{\bm X}'\overline{\bm X} + \underbrace{\frac{\omega}{1-\omega} \overline{\bm X}' \left[\begin{pmatrix} \bm I_T & \bm 0 \\ \bm 0 & \bm 0\end{pmatrix} - \begin{pmatrix} \bm \Psi_0 & \bm 0 \\ \bm 0 & \bm 0\end{pmatrix}\right] \overline{\bm X}}_{\underline{\bm V}^{-1}_{\Psi_0}}\right)^{-1}.
\end{equation*}
This equation implies that $\underline{\bm V} = \left(\underline{\bm X}'\underline{\bm X} + \underline{\bm V}^{-1}_{\Psi_0}\right)^{-1}$. Adding the Minnesota prior means that the result in (\ref{eq: convex}) no longer holds exactly. Intuitively speaking, if $\vartheta$ is set too tight, the Minnesota-type prior overrules the subspace shrinkage prior. However, for reasonable values of $\vartheta$ the following will hold approximately:
\begin{equation*}
\mathbb{E}(\bm X \bm A | \bm Y, \vartheta, \omega) \approx  \omega \bm \Phi_0 \bm Y + (1-\omega) \bm X (\overline{\bm X}' \overline{\bm X})^{-1}\overline{\bm X}' \overline{\bm Y}.
\end{equation*}
This result states that the posterior mean of the regression function is a convex combination of the (OLS) fit of a PCA regression and the posterior mean based on a Minnesota prior VAR.

To investigate the accuracy of this approximation for different values of values of $\vartheta$ we can  compute the average squared approximation error:
\begin{equation}
\Xi(\theta, \omega) = \frac{1}{M} \sum_{j=1}^M T^{-1}||\bm X \overline{\bm A}_{j} - \omega \bm \Phi_0 \bm Y_j - (1-\omega) \bm X (\overline{\bm X}' \overline{\bm X})^{-1}\overline{\bm X}' \overline{\bm Y}_j||^2,
\end{equation}
with $\overline{\bm A}_{j}, \bm Y_j$ and $\overline{\bm Y}_j$ denoting the $j^{th}$ column of the corresponding matrix and $|| \bullet ||$ denotes the Euclidean norm of a vector. This approximation error quickly approaches zero if $\vartheta$ becomes moderately large. If $\omega \approx 0$, the approximation error also vanishes since then we obtain the Minnesota prior BVAR estimate. The interaction between $\vartheta$ and $\omega$ in determining $\Xi(\vartheta, \omega)$ is highly non-linear. The key point to take away is that if $\vartheta$ approaches zero faster than $\omega$ approaches one, the standard Minnesota prior dominates the subspace prior. 

\begin{figure}
\begin{minipage}[t]{.5\textwidth}
\centering $M=30$
\end{minipage}
\begin{minipage}[t]{.5\textwidth}
\centering $M=120$
\end{minipage}\\
\begin{minipage}[t]{1\textwidth}
\centering $T=250$
\end{minipage}
\begin{minipage}[t]{0.5\textwidth}
\includegraphics[scale=0.42]{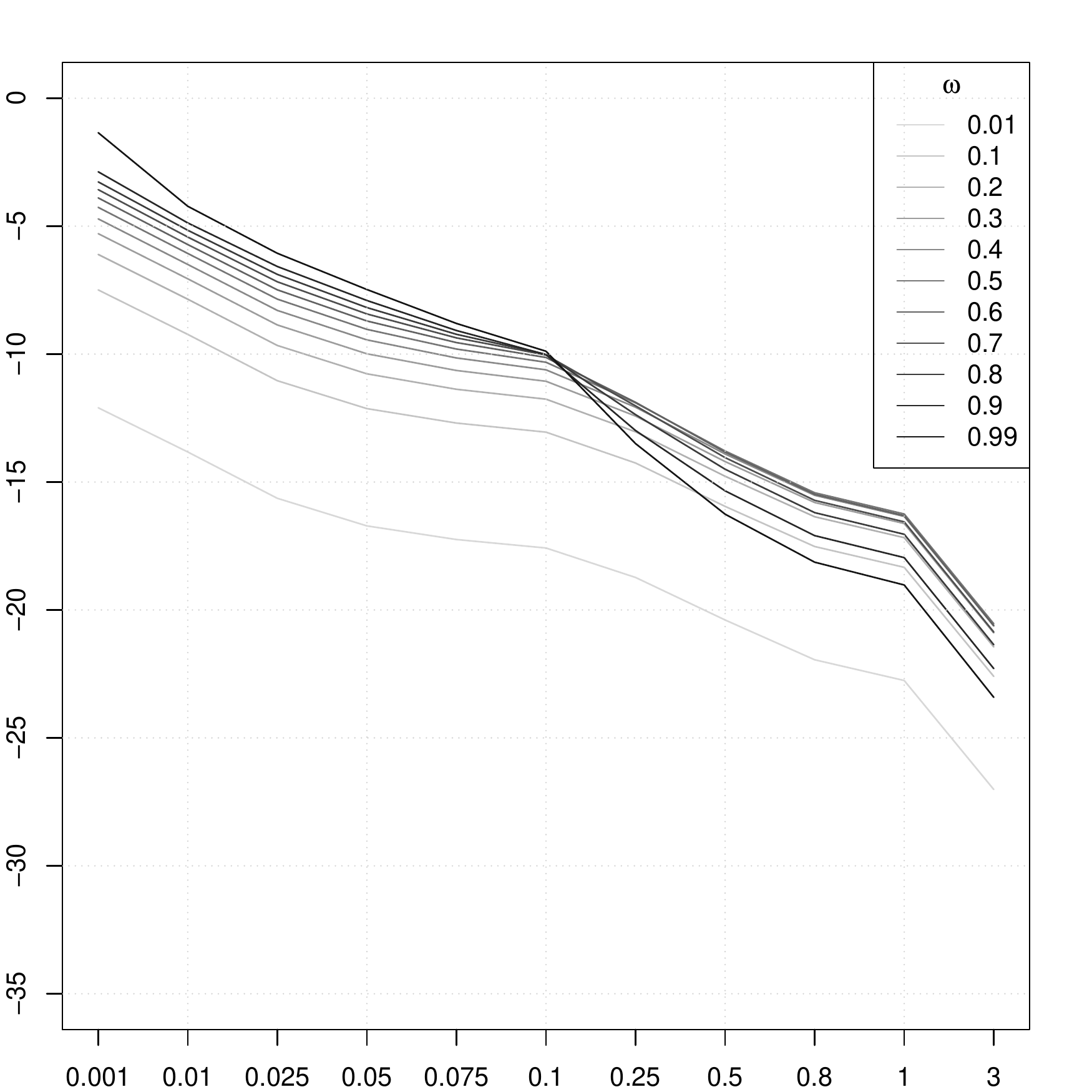}
\end{minipage}
\begin{minipage}[t]{0.5\textwidth}
\includegraphics[scale=0.42]{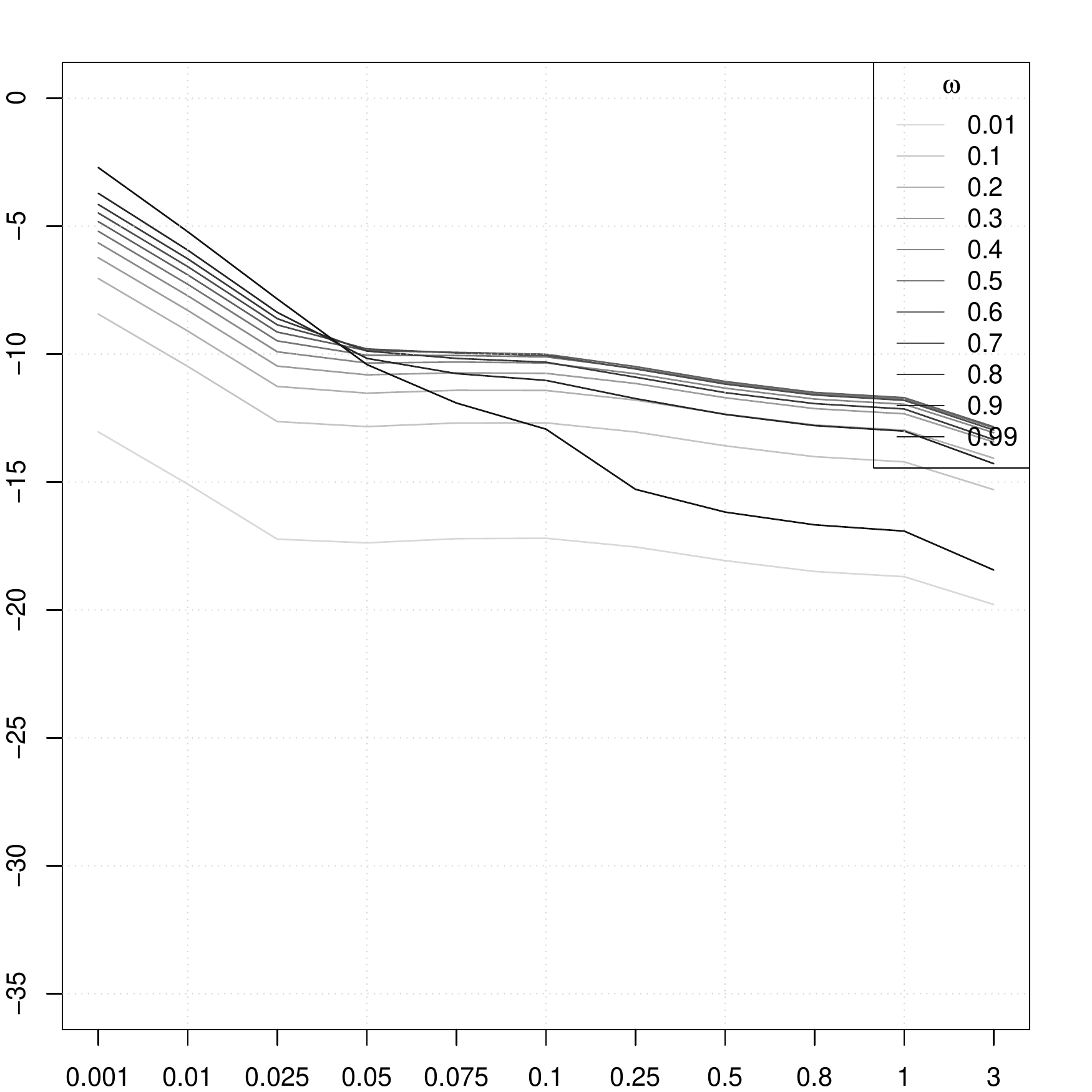}
\end{minipage}
\begin{minipage}[t]{1\textwidth}
\centering $T=350$
\end{minipage}
\begin{minipage}[t]{0.5\textwidth}
\includegraphics[scale=0.42]{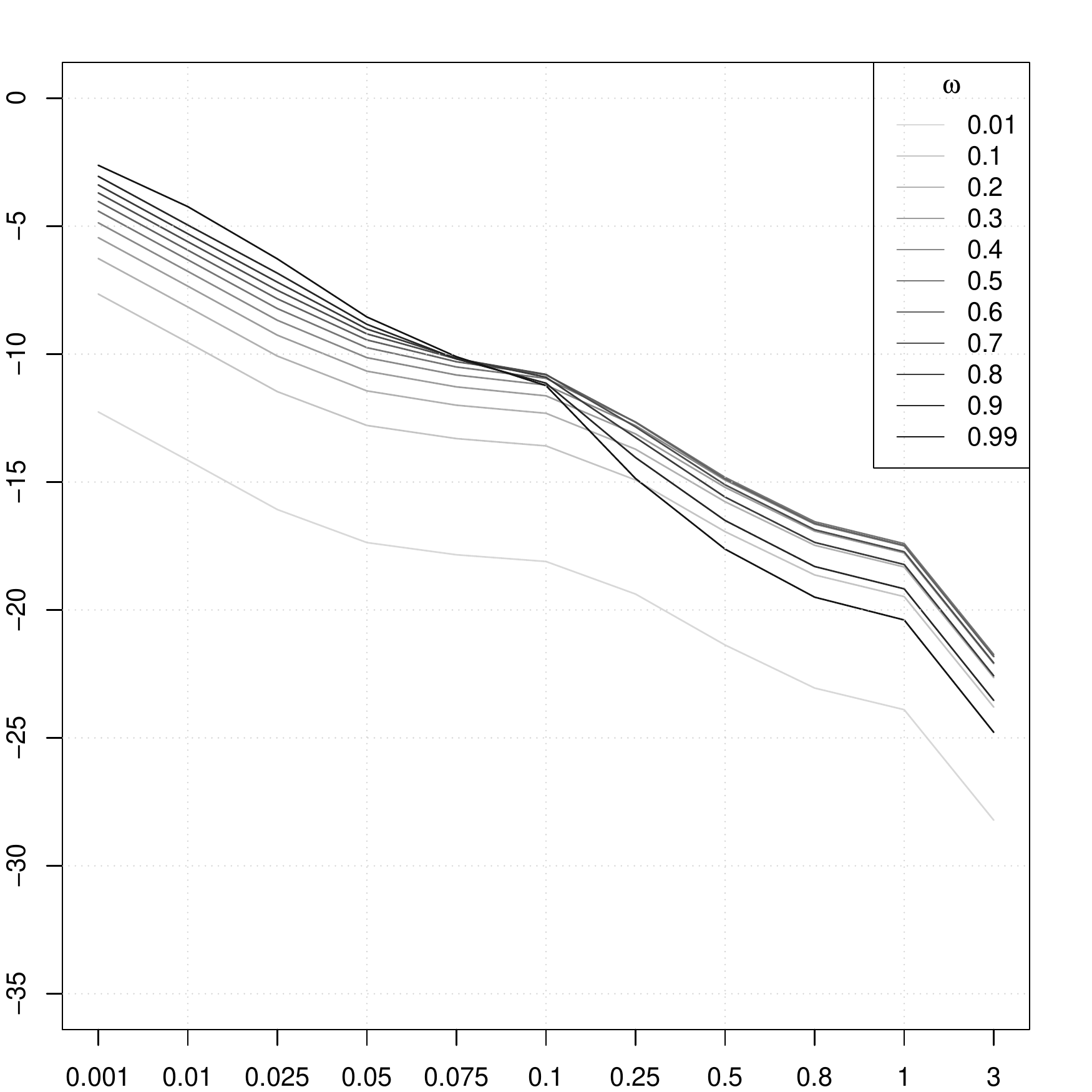}
\end{minipage}
\begin{minipage}[t]{0.5\textwidth}
\includegraphics[scale=0.42]{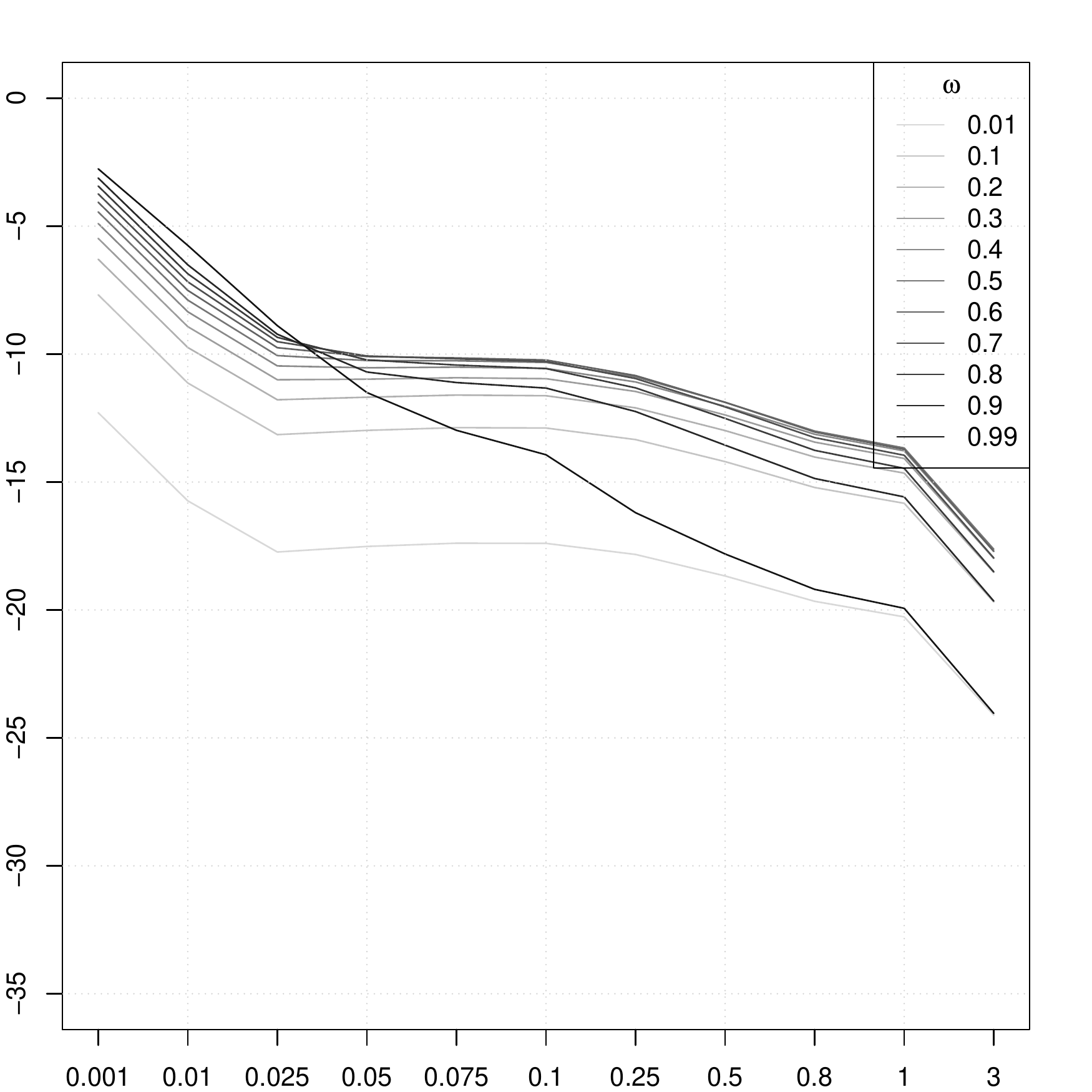}
\end{minipage}
\begin{minipage}[t]{1\textwidth}
\centering $T=500$
\end{minipage}
\begin{minipage}[t]{0.5\textwidth}
\includegraphics[scale=0.42]{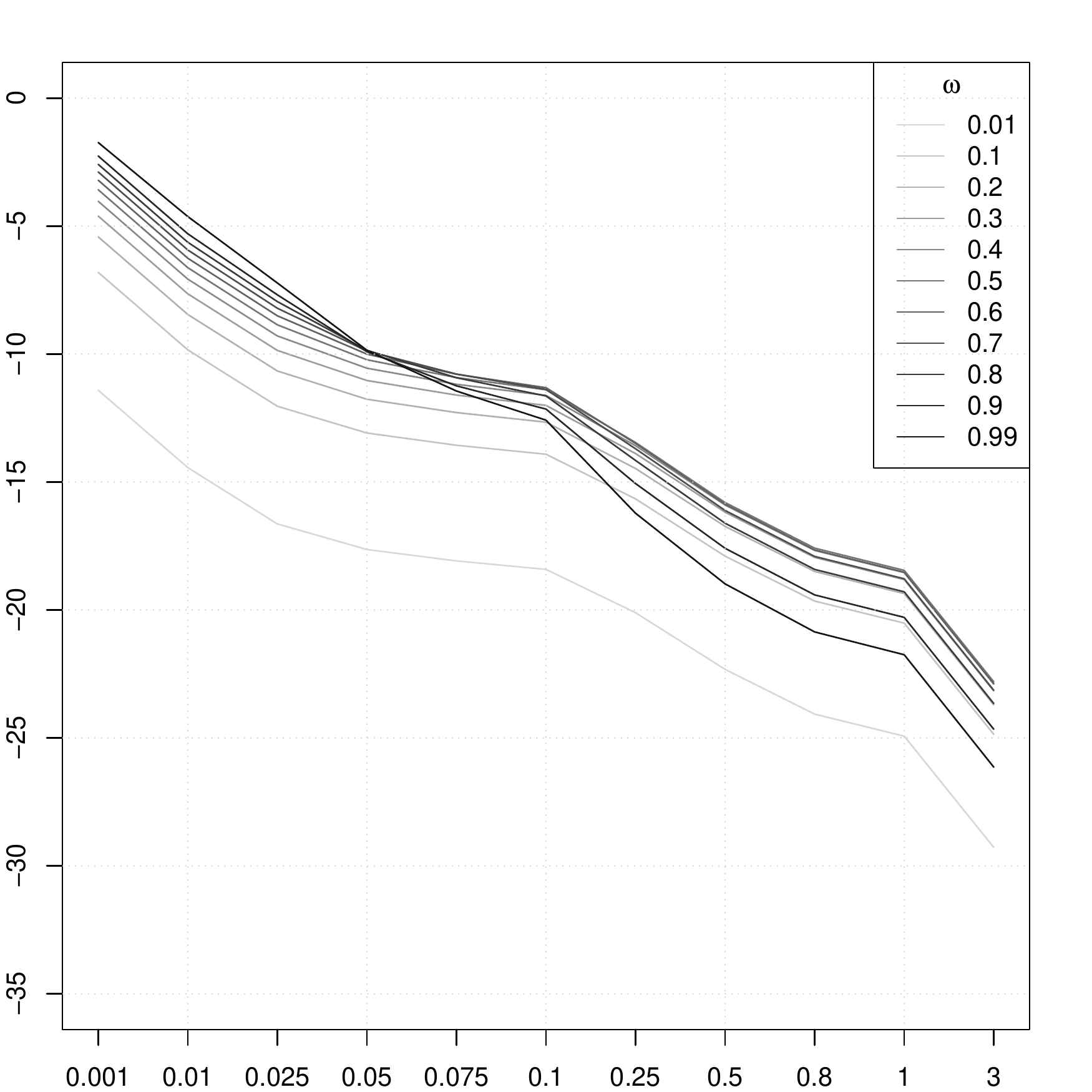}
\end{minipage}
\begin{minipage}[t]{0.5\textwidth}
\includegraphics[scale=0.42]{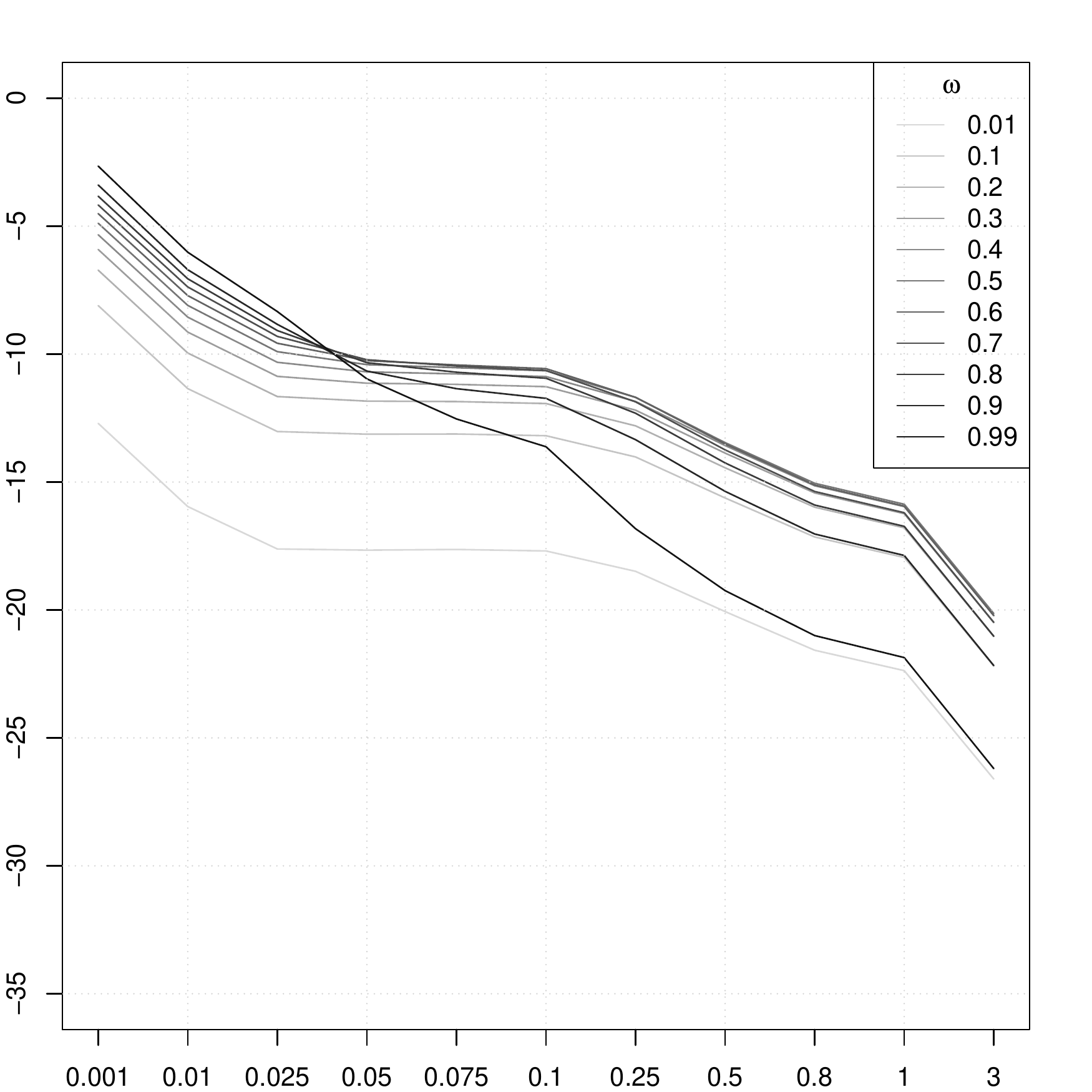}
\end{minipage}

\caption{Log Squared Approximation Error for different DGPs and $q=3$}
\label{approx_error}
\end{figure}

These points are illustrated in Figure \ref{approx_error} which plots the approximation error for different values of $\vartheta$ and $\omega$ using data sets simulated from different data generating processes (DGPs)  for different values of $M$ and $T$. The DGP is a dynamic factor model with $q=3$ and the factors evolving according to a multivariate random walk with a full error variance-covariance matrix.\footnote{For more details on the DGP, see Section \ref{sec: simulation}.}

Figure \ref{approx_error} suggests that $\omega$ does not have a large effect on the approximation, but that $\vartheta$ does. In particular, for values of $\vartheta>0.1$, the log approximation error is less than $-8$ for all the different values of $\omega$. In the next section we will specify a prior on $\vartheta$ which allocates substantial mass to this region. It is worth stressing that even if $\vartheta$ is smaller than this, our prior is still a valid prior combining the Minnesota prior with the subspace prior, it is just that the posterior mean that results will deviate more from being a linear combination of a posterior mean using the Minnesota prior and a PC regression.

\subsection{Selecting the number of factors and  estimating the hyperparameters} \label{sec: factorestimation}

Our prior depends on a choice for the number of factors ($q$), the weight attached to the VAR relative to the PC regression ($\omega$) and the degree of shrinkage in the Minnesota prior ($\vartheta$).
The posterior for these is:
\begin{equation*}
p(q, \omega, \vartheta | \bm Y) \propto p(\bm Y | q, \omega, \vartheta)~p( q, \omega, \vartheta).
\end{equation*}
Estimation is straightforward since the natural conjugate prior leads to an analytical form for the marginal likelihood:
\begin{equation}
p(\bm Y | q, \omega, \vartheta) \propto \left( |\underline{\bm V}| ~ |\overline{\bm V}| \right)^{-\frac{M}{2}} ~ |\overline{\bm S}|^{-\frac{T+M+\underline{v}}{2}}, 
\label{eq: marglike}
\end{equation}
where 
 $\overline{\bm S} = \underline{\bm S} + \bm Y' \bm Y + \underline{\bm A}' \underline{\bm V}^{-1} \underline{\bm A} - \overline{\bm A}' \overline{\bm V}^{-1} \overline{\bm A} $ is the posterior scaling matrix of the inverse Wishart posterior of $\bm \Sigma$. This can be multiplied by the prior to produce the posterior. We define discrete grids for $\omega$, $\vartheta$ and $q$ and evaluate the posterior at points in the grids. We can then do Monte Carlo integration by sampling from the Multinomial distributions that arise. Hence, our predictive densities reflect uncertainty in these parameters. This is similar to a strategy suggested in \cite{GiannoneLenzaPrimiceri2015RESTAT} for the Minnesota prior VAR but, as detailed below, avoids carrying out complex matrix operations during posterior simulation and thus offers substantial computational gains (at the cost of approximating a continuous posterior distribution using a discrete one).

It remains to specify the priors. For $\vartheta$ we follow suggestions in \cite{GiannoneLenzaPrimiceri2015RESTAT} and use a Gamma prior which we set to have mode $0.2$ and standard deviation $0.4$. This value implies that the approximation error in (\ref{approx_error}) is extremely small and  the posterior mean of the model fit can be safely interpreted as a convex combination between the BVAR  and the DFM  fit.

For $\omega$ we use a Beta prior: $\mathcal{B}(c_0, c_1)$. In our empirical work  we consider two ways of specifying the hyperparameters  $c_0$ and $c_1$. The first sets $c_0 = c_1 =1$, yielding a non-informative uniform prior on $\omega$. The second prior sets $c_0 = \mathfrak{c}_0\times M$ and $c_1 = \mathfrak{c}_1 \times M$, with $\mathfrak{c}_0, \mathfrak{c}_1$ being scalars greater than zero. This choice implies that the prior mean on $\omega$ is equal to $\mathfrak{c}_0/(\mathfrak{c}0+ \mathfrak{c}_1)$ and the prior variance equals $(\mathfrak{c}_0 ~ \mathfrak{c}_1)/((\mathfrak{c}_0+ \mathfrak{c}_1)^2~(M (\mathfrak{c}_0+\mathfrak{c}_1)+1)$.  In our empirical work we set $\mathfrak{c}_0 = 8$ and $\mathfrak{c}_1 = 6$, yielding a prior mean on $\omega$ of around $0.6$ and thus placing considerable mass on the factor model restrictions while the prior variance decreases in $M$. In large dimensions, this choice increasingly forces the model towards the factor restrictions but still provides sufficient flexibility for individual time series to exhibit VAR dynamics.

We assume a discrete uniform prior on $q$:
\begin{equation*}
    q \sim \mathcal{U}(1, q_0), 
\end{equation*}
which implies that all values up to $q_0$ (which denotes some integer smaller than $K$ set by the researcher) are a-priori equally likely. Other choices which utilize sample information (such as the eigenvalues of $\bm X$) are in principle possible. 

Two hyperparameters remain to be chosen. If we use a flat prior in combination with a subspace prior, we set $\underline{v}=M+2$ and $\underline{\bm S} = \frac{1}{100} \bm I_M$. If we use a Minnesota prior we set $\underline{v}$ equal to the number of rows of $\underline{\bm Y}$ and $\underline{\bm S} =(\underline{\bm Y} - \underline{\bm X} \underline{\bm A})'(\underline{\bm Y} - \underline{\bm X} \underline{\bm A})$ (see \cite{kadiyala1997numerical}). 

 It is worth noting that, in Bayesian factor analysis, selecting the number of latent factors is a difficult task. Bayesian solutions include using reversible jump MCMC algorithms which treat the number of factors as an unknown quantity (see, e.g., \cite{LWfactor} and \cite{Lopes_FS}). Another strand of the literature estimates an overfitting factor model and applies Bayesian shrinkage priors to force the columns of the factor loadings matrix associated with irrelevant factors to zero, see \cite{BD2011}. However, the following sub-section investigates, using simulated data, the simple and computationally efficient approach given here. These simulations show that, even under a uniform prior, our approach selects the true number of factors successfully. 
 
 It is also worth noting that, conditional on $\omega$ and $\vartheta$, all quantities used in the Monte Carlo sampling of $q$  can be pre-computed and thus estimation of huge models (i.e., with $M > 100$) is feasible. This requires specifying a grid for $\omega$, $\theta$ and $q$. In all our empirical work we set the grid for $q \in \{1, \dots, \text{min}(10, \mathcal{L}^*)\}$ with $\mathcal{L}^*$ denoting the Ledermann bound.\footnote{The Ledermann bound is defined as the solution $\mathcal{L}^*$ to the equation $(M-\mathcal{L}^*)^2 \ge M+\mathcal{L}^*$.} The grid on $\omega$ is specified to go from $0.01$ to $0.99$ with a step-size of $0.05$. Finally, the grid on $\vartheta$ is $\{0.001, 0.01, 0.025, 0.05, 0.10, 0.20, 0.3, 0.4, 0.5, 2, 3, 4, 5\}$.
 
 Evaluating marginal likelihoods can be challenging in very large models and they depend on the prior. Accordingly, in our empirical work (which involves forecasting three variables of interest), we also investigate an alternative way of choosing $q$, $\omega$ and $\vartheta$. This is to use the Bayesian Information Criterion (BIC) for the three variables of interest to choose them.

\subsection{Simulated data exercise on selecting the number of factors}\label{sec: simulation}
In this sub-section we investigate whether our approach successfully detects the correct number of factors by means of synthetic data. To analyze how estimation accuracy changes across different datasets, we consider DGPs that vary along the number of variables ($M$) as well as the number of factors. The  DGP is a dynamic factor model given by:
\begin{align*}
  \bm y_t = \bm \Lambda \bm f_{t-1} + \bm \epsilon_t,\quad \bm \epsilon_t \sim \mathcal{N}(\bm 0_M, \bm \Sigma),
\end{align*}
with $\bm f_t$ evolving according to a multivariate random walk with full state-innovation variance $\bm \Omega$, an initial state $\bm f_0 = \bm 0_q$ and $\bm \Sigma = \bm \Lambda \bm \Omega \bm \Lambda' + \bm W$ being a full matrix with $\bm W$ denoting a diagonal matrix of measurement error variances. Notice that this is a standard dynamic factor model which is rewritten by plugging the random walk state equation into an observation equation which typically includes the contemporaneous factors and uncorrelated measurement errors.  

In all our simulations we assume that $\lambda_{ij}$, the $(i,j)^{th}$ element of $\bm \Lambda$, is drawn from a normal distribution with zero mean and variance $0.1^2$ if $i \neq j$ or set equal to unity if $i = j$. Instead of specifying $\bm \Omega$ we obtain the lower Cholesky factor  of $\bm \Sigma$, $\bm A^{-1}_0$,   by simulating the off-diagonal elements from a Gaussian distribution with zero mean and variance $0.1^2$ and the main diagonal elements are set equal to $0.1$.

We simulate $T=500$ observations from small ($M=10$), moderate ($M=60$) and large ($M=120$) datasets.  For each of these, we vary the number of factors $q \in \{1, 3, 6, 8\}$. All simulations are repeated $100$ times and, in \autoref{tab:simulation}, we report averages of posterior medians across these replications.

It can be seen that all of the versions of our subVAR prior are doing a good job of choosing the correct number of factors. It is mainly in the least parsimonious cases (i.e. DGPs with $M=120$ and $q=8$) where it is slightly underestimating the number of factors. But this is due to the large VAR providing some of the fit, leaving less for the DFM to explain. In the context of these very large models, slight over-shrinkage is better than the over-fitting which would have occurred if the prior had failed to shrink enough.

\begin{table}[t]
\caption{Simulation results for differing values of $q$ and $M$. Averages across 100 replications from the DGP} \label{tab:simulation}
\centering
\scalebox{0.82}{
\begin{tabular}{lrrrrrrrrrrrrrrrr}
  \toprule
  \multicolumn{4}{r}{subVAR-Minn0}& \multicolumn{4}{r}{subVAR-Minn1} & \multicolumn{4}{r}{subVAR-Flat0} & \multicolumn{4}{r}{subVAR-Flat1}\\
$q=$ & 1 & 3 & 6 & 8 & 1 & 3 & 6 & 8 & 1 & 3 & 6 & 8 & 1 & 3 & 6 & 8 \\ 
  \midrule
  \multicolumn{16}{l}{Posterior mean of $q$}\\
$M=10$ & 1.32 & 3.03 & 5.87 & 2.54 & 1.29 & 3.11 & 6.00 & 8.00 & 1.32 & 3.04 & 5.90 & 2.53 & 1.28 & 3.06 & 6.00 & 8.00 \\ 
  $M=60$ & 1.00 & 2.96 & 5.23 & 6.01 & 1.03 & 2.95 & 5.34 & 6.24 & 1.00 & 2.95 & 5.26 & 6.01 & 1.02 & 2.99 & 5.28 & 6.22 \\ 
  $M=120$ & 1.00 & 2.58 & 3.92 & 4.54 & 1.00 & 2.67 & 3.92 & 4.59 & 1.00 & 2.57 & 3.84 & 4.50 & 1.00 & 2.63 & 3.95 & 4.61 \\ 
   \bottomrule
\end{tabular}
}
  \smallskip
\begin{minipage}{\linewidth}\small
\tiny \textbf{Notes}: subVAR denotes the VAR coupled with the subspace shrinkage prior, Minn is the combination between subspace and Minnesota shrinkage while flat is the subspace shrinkage prior without additional shrinkage. The $0$ and $1$ attached to the respective label indicate a flat  ($0$) or informative ($1$) prior on $\omega$. Each number is based on computing the mean of posterior medians across 100 replications from the respective DGPs. For $q$, we use the posterior median as our point estimate while for $\omega$ we use the posterior mean.
\end{minipage}
\end{table}

\section{Forecasting Using US Macroeconomic Data}\label{sec: appl}
\subsection{Data}
We use a large set of $166$ quarterly macroeconomic variables taken from the St. Louis Fed's FRED data base (\href{https://fred.stlouisfed.org}{fred.stlouisfed.org}) and discussed in \cite{mccracken2020fred}. These are listed in the appendix in Table \ref{tab: dataset}. Variables are transformed to stationarity following recommendations there. 
Our forecasting results focus on three variables of interest: GDP growth (based on real GDP growth, GDPC1), the Fed Funds rate (FEDFUNDS) and inflation (based on the consumer price index, CPIAUCSL). 

The data runs from 1960:Q1 to 2020:Q3 and in our forecasting exercise, the evaluation period is from 1990:Q3-2020:Q3. We adopt a recursive forecasting design. We use the initial estimation period (1960:Q1 to 1990:Q2) to produce one- and iterated four-quarter-ahead forecast distributions for 1990:Q3 and 1991:Q2, respectively. After obtaining these, we expand the initial estimation period by one observation until we reach the end of the sample.

\subsection{Models}
We present results for four models involving subspace priors (acronym subVAR). These involve two priors for the VAR coefficients: the non-informative one (flat) and the Minnesota prior (Minn). There are also two priors for $\omega$: one non-informative (flat) and one informative. These are indicated by adding a $0$ (flat) and $1$ (tight) to the relevant labels of the VAR coefficient prior.

For each of these four models, we present results for data sets of four different sizes: small (S, $12$ variables), medium (M, $22$ variables), large (L, $78$ variables) and extra large (XL, $166$ variables). Table \ref{tab: dataset} lists which variable belongs in which category. 

For comparison we also present results for Minnesota prior VARs (implemented by setting $\omega=0$ in the subVAR-Minn) and  factor models (labeled DFM). The factor model is a FAVAR for the three variables of interest and the factors estimated by extracting the PCs from the remaining time series within a given model size. The number of PCs is chosen by retaining the PCs with standard deviations greater than unity. It is estimated using a relatively non-informative Minnesota prior.  The lag length in all models is set to two. 

\subsection{Summary of forecasting results}
We begin by summarizing the results of our pseudo-out-of-sample forecasting exercise in Table \ref{tab: forecast}. This table contains Root Mean Squares Forecast Errors (RMSFEs) and averages (over time) of log predictive likelihoods (LPLs) for our three variables of interest and for two different forecast horizons.  The RMSFEs are ratios between the RMSFEs of a given model and the Minnesota VAR while the LPLs are differences between a given model and the Minnesota VAR (both for a given model size).

Before discussing our subVAR models, consider the comparison between the Minnesota prior VAR and the factor model. For some variables, forecast horizons and model sizes, the VAR yields more precise forecasts. That is, for the interest rate it consistently forecasts better and for inflation and GDP growth for larger models at longer horizons, its forecasts tend to be more precise than the ones of the DFM. But for other cases the factor model outperforms the Bayesian VAR.  This result raises the possibility that an approach such as ours, which combines the two, could lead to better overall forecast performance than either the BVAR or DFM individually and, with several exceptions discussed below, this is what we find. 

Consider first the most informative subVAR model which uses the Minnesota prior on the VAR coefficients and the informative prior on $\omega$ (Minn1). With some exceptions, this model is yielding forecasts which are better than the predictions produced by the BVAR and are often the most precise ones. The main exceptions are the one-year-ahead GDP growth predictions which are marginally worse than the the BVAR benchmark. However, this is a case where the DFM and some of the less informative subVAR approaches are forecasting substantially worse than the BVAR. 

Consider now the second most informative subVAR approach (Minn0) which retains the Minnesota prior for the VAR coefficients, but uses a non-informative prior on $\omega$. Its forecasts are comparable to those of Minn1, but overall are  slightly worse. But clearly results are robust to the prior on $\omega$. Both of these Minnesota prior subVAR approaches are forecasting well most of the time and even the few exceptions reveal only slight deterioration in forecast performance relative to the BVAR benchmark.

Using a non-informative prior for the VAR coefficients, however, goes wrong in some cases in larger models. In one sense, this is unsurprising. Non-informative priors work poorly in large VARs since they suffer from severe over-fitting problems. It might have been possible that these would have been corrected by adding the subspace prior shrinking towards the factor model. But clearly this effect is not strong enough in the L and XL models to counter-balance the over-fitting problem. One might have hoped that estimates of $\omega$ would have been pulled towards $1$ in these cases, leading to results similar to the DFM but (unless we use an extremely dogmatic prior on $\omega$) this is not happening in the larger models. Similar to the results based on synthetic data, this is because the larger (unrestricted) VARs explain the majority of variation in the data and leave little variation to explain for the factor model, yielding posterior estimates of $\omega$ close to zero. However, the subVAR-Flat models are performing well for our small and medium-sized models and for the one-quarter-ahead forecast horizon for the larger models. And it is only the iterated one-year-ahead forecasts that are deteriorating. This suggests that this approach might be found useful by researchers working with VARs up to a dimension of approximately $20$ who wish to avoid the use of standard BVAR priors such as the Minnesota prior, particularly if the focus is on short-term forecasts. Such a researcher may also wish to avoid using marginal likelihoods since they are prior-dependent. For them, it is also interesting to note that using the BIC to estimate the prior hyperparameters also works roughly as well as using marginal likelihoods. 

\begin{table}[h!]
\caption{Forecasting results across focus variables, models and forecast horizons}\label{tab: forecast}
\centering
\scalebox{0.65}{
\begin{tabular}{lrrrrrrrrrrrrrrrrrr}
  \toprule
    & \multicolumn{6}{c}{\textbf{FEDFUNDS}} & \multicolumn{6}{c}{\textbf{CPIAUCSL}} & \multicolumn{6}{c}{\textbf{GDPC1}}\\
     & \multicolumn{4}{c}{SubVAR} &  DFM&  BVAR  & \multicolumn{4}{c}{SubVAR} & DFM& Minn & \multicolumn{4}{c}{SubVAR} & DFM & Minn  \\
      \cmidrule(lr){2-7}\cmidrule(lr){8-13}\cmidrule(lr){14-19}
 &  Minn0 & Minn1  &  Flat0 & Flat1 &  &  &  Minn0& Minn1 &  Flat0 & Flat1&  &  & Minn0& Minn1 &  Flat0 & Flat1&  &  \\ 
  \midrule
          \multicolumn{19}{l}{\textbf{Marginal Likelihood to estimate $q$, $\omega$ and $\vartheta$}}\\
        \multicolumn{19}{l}{Relative RMSEs}\\
            \multicolumn{19}{l}{\textit{One-quarter-ahead}}\\
  S & 1.02 & 0.95 & 1.03 & 0.96 & 1.07 & 0.58 & 0.99 & 0.98 & 0.99 & 0.98 & 1.01 & 1.21 & 0.99 & 0.99 & 1.00 & 0.99 & 1.00 & 1.15 \\ 
  M & 1.04 & 0.93 & 1.12 & 1.05 & 1.18 & 0.64 & 0.99 & 0.97 & 1.01 & 1.00 & 0.98 & 1.26 & 1.00 & 1.00 & 1.04 & 1.02 & 1.02 & 1.14 \\ 
  L & 0.96 & 0.94 & 1.89 & 1.79 & 1.37 & 0.74 & 0.98 & 0.98 & 1.33 & 1.30 & 0.96 & 1.32 & 0.99 & 0.99 & 1.41 & 1.38 & 0.92 & 1.19 \\ 
  XL & 0.96 & 0.91 & 1.48 & 1.29 & 1.37 & 0.74 & 0.98 & 0.96 & 1.08 & 1.04 & 1.05 & 1.35 & 0.99 & 0.97 & 1.11 & 1.07 & 0.93 & 1.21 \\ 
   \multicolumn{19}{l}{\textit{One-year-ahead}}\\
  S & 1.00 & 1.01 & 1.03 & 1.01 & 1.03 & 0.49 & 1.01 & 1.01 & 1.00 & 1.00 & 1.03 & 1.14 & 1.00 & 1.00 & 0.99 & 1.00 & 1.00 & 1.12 \\ 
  M & 1.04 & 0.95 & 1.19 & 1.11 & 0.95 & 0.53 & 1.01 & 1.00 & 1.02 & 1.01 & 1.06 & 1.14 & 1.00 & 1.00 & 1.01 & 1.00 & 1.01 & 1.12 \\ 
  L & 0.90 & 0.88 & 6.16 & 9.47 & 1.48 & 0.69 & 0.99 & 0.98 & 2.72 & 3.74 & 1.03 & 1.18 & 1.00 & 1.00 & 3.10 & 5.54 & 1.07 & 1.12 \\ 
  XL & 0.87 & 0.71 &  $>10$ & $>10$  & 1.42 & 0.85 & 0.99 & 0.97 & $>10$ & $>10$ & 1.06 & 1.19 & 1.00 & 0.97 & $>10$ & $>10$ & 1.03 & 1.13 \\ 
          \multicolumn{19}{l}{Average  LPLs}\\
          \multicolumn{19}{l}{\textit{One-quarter-ahead}}\\
  S & -0.01 & 0.05 & 0.01 & 0.06 & -0.52 & -0.91 & 0.05 & 0.11 & 0.05 & 0.13 & 0.32 & -2.00 & 0.02 & 0.06 & 0.05 & 0.05 & 0.00 & -1.67 \\ 
  M & -0.01 & 0.05 & -0.05 & -0.01 & -0.63 & -0.87 & 0.01 & 0.20 & 0.07 & 0.18 & 0.75 & -2.36 & 0.01 & 0.01 & -0.03 & -0.02 & 0.10 & -1.66 \\ 
  L & 0.11 & 0.14 & -0.63 & -0.46 & -0.67 & -1.13 & 0.26 & 0.39 & 1.37 & 1.53 & 2.06 & -3.76 & -0.04 & 0.05 & -0.47 & -0.45 & 0.10 & -1.69 \\ 
  XL & 0.84 & 1.36 & 0.84 & 0.56 & 0.62 & -2.49 & 2.12 & 3.28 & 4.61 & 4.29 & 4.65 & -6.44 & 0.28 & 0.36 & 0.15 & -0.21 & 0.71 & -2.36 \\ 
  \multicolumn{19}{l}{\textit{One-year-ahead}}\\
  S & -0.07 & 0.05 & -0.26 & -0.16 & -0.47 & -1.08 & 0.02 & -0.00 & -0.04 & 0.00 & 0.04 & -1.64 & -0.02 & 0.02 & -0.07 & -0.05 & 0.02 & -1.63 \\ 
  M & -0.07 & 0.04 & -0.49 & -0.40 & -0.62 & -1.07 & 0.06 & 0.01 & 0.05 & 0.05 & 0.05 & -1.75 & -0.06 & -0.05 & -0.20 & -0.15 & -0.05 & -1.60 \\ 
  L & 0.03 & 0.06 & -3.64 & -3.93 & -2.14 & -1.57 & 0.03 & 0.03 & -3.45 & -3.73 & -1.59 & -1.58 & -0.02 & -0.03 & -3.74 & -4.04 & -1.29 & -1.66 \\ 
  XL & -0.12 & -0.31 & -6.61 & -8.18 & -2.01 & -1.45 & -0.09 & -0.28 & -6.60 & -8.23 & -1.39 & -1.56 & -0.19 & -0.49 & -7.03 & -8.63 & -1.25 & -1.55 \\ \midrule
      \multicolumn{19}{l}{\textbf{BIC for the three focus variables used to estimate $q$, $\omega$ and $\vartheta$}}\\
              \multicolumn{19}{l}{Relative RMSEs}\\
            \multicolumn{19}{l}{\textit{One-quarter-ahead}}\\
  S & 0.96 & 0.91 & 0.97 & 0.88 & 0.96 & 0.66 & 0.99 & 0.95 & 0.99 & 0.95 & 0.98 & 1.24 & 1.00 & 0.99 & 1.00 & 0.99 & 0.99 & 1.15 \\ 
  M & 0.92 & 0.74 & 0.94 & 0.74 & 0.88 & 0.85 & 0.97 & 0.89 & 0.98 & 0.89 & 0.90 & 1.37 & 0.99 & 0.93 & 1.00 & 0.95 & 0.97 & 1.21 \\ 
  L & 0.93 & 0.82 & 1.81 & 1.15 & 1.47 & 0.71 & 0.98 & 0.92 & 1.30 & 1.07 & 0.97 & 1.30 & 0.98 & 0.96 & 1.35 & 1.09 & 0.92 & 1.18 \\ 
  XL & 0.94 & 0.84 & 1.43 & 1.03 & 1.45 & 0.71 & 0.97 & 0.91 & 1.05 & 0.96 & 1.06 & 1.34 & 0.99 & 0.94 & 1.08 & 0.99 & 0.94 & 1.20 \\ 
     \multicolumn{19}{l}{\textit{One-year-ahead}}\\
  S & 0.98 & 0.94 & 1.00 & 0.95 & 0.93 & 0.52 & 1.01 & 0.99 & 1.00 & 0.99 & 1.01 & 1.16 & 1.01 & 1.01 & 1.01 & 1.02 & 1.01 & 1.11 \\ 
  M & 0.84 & 0.56 & 0.89 & 0.56 & 0.56 & 0.94 & 0.97 & 0.92 & 0.97 & 0.93 & 0.97 & 1.25 & 0.97 & 0.94 & 0.97 & 0.93 & 0.95 & 1.20 \\ 
  L & 0.87 & 0.75 & 8.15 & $>10$ & 1.63 & 0.64 & 0.99 & 0.99 & 3.77 & $>10$ & 1.06 & 1.15 & 0.99 & 0.99 & 5.31 & $>10$ & 1.09 & 1.12 \\ 
  XL & 0.84 & 0.65 & $>10$ & $>10$ & 1.66 & 0.76 & 1.00 & 0.98 & $>10$ & $>10$ & 1.07 & 1.17 & 0.98 & 0.98 &$>10$  & $>10$ & 1.04 & 1.11 \\ 
       \multicolumn{19}{l}{Average  LPLs}\\
          \multicolumn{19}{l}{\textit{One-quarter-ahead}}\\
  S & 0.05 & 0.09 & 0.05 & 0.12 & -0.42 & -1.01 & 0.05 & 0.19 & 0.09 & 0.22 & 0.43 & -2.09 & 0.03 & 0.06 & 0.02 & 0.05 & 0.07 & -1.67 \\ 
  M & 0.07 & 0.20 & 0.06 & 0.21 & -0.45 & -1.05 & 0.18 & 0.63 & 0.13 & 0.60 & 1.23 & -2.83 & -0.01 & 0.05 & 0.01 & 0.07 & 0.15 & -1.70 \\ 
  L & 0.14 & 0.22 & -0.48 & -0.31 & -0.76 & -1.04 & 0.39 & 0.81 & 1.20 & 1.63 & 1.69 & -3.39 & 0.13 & 0.17 & -0.33 & -0.27 & 0.17 & -1.78 \\ 
  XL & 0.63 & 1.02 & -0.06 & -0.55 & -0.06 & -1.81 & 1.53 & 2.45 & 2.91 & 2.47 & 3.19 & -4.99 & 0.16 & 0.31 & -0.34 & -0.78 & 0.48 & -2.13 \\ 
    \multicolumn{19}{l}{\textit{One-year-ahead}}\\
  S & 0.06 & 0.29 & 0.00 & 0.26 & -0.14 & -1.41 & 0.02 & 0.06 & 0.01 & 0.06 & 0.11 & -1.69 & -0.01 & 0.06 & -0.04 & 0.00 & 0.07 & -1.67 \\ 
  M & 0.09 & 0.57 & 0.06 & 0.54 & 0.03 & -1.72 & 0.04 & -0.02 & -0.01 & 0.04 & 0.03 & -1.73 & 0.05 & 0.20 & 0.04 & 0.19 & 0.20 & -1.83 \\ 
  L & 0.06 & 0.25 & -4.22 & -5.04 & -2.21 & -1.50 & -0.00 & -0.03 & -3.98 & -4.77 & -1.63 & -1.55 & -0.02 & -0.05 & -4.29 & -5.15 & -1.32 & -1.64 \\ 
  XL & -0.20 & -0.73 & -8.11 & -9.95 & -1.98 & -1.45 & -0.20 & -0.74 & -8.15 & -9.97 & -1.37 & -1.57 & -0.27 & -0.96 & -8.44 & -7.59 & -1.13 & -1.65 \\ 
   \bottomrule
\end{tabular}}
   \smallskip
\begin{minipage}{\linewidth}\small
\tiny \textbf{Notes}: subVAR denotes the VAR coupled with the subspace shrinkage prior, Minn is the combination between subspace and Minnesota shrinkage while flat is the subspace shrinkage prior without additional shrinkage. The $0$ and $1$ attached to the respective label indicate a flat  ($0$) or informative ($1$) prior on $\omega$. DFM is a VAR in the three focus variables augmented with principal components and BVAR refers to a Minnesota VAR. The upper part of the table shows relative root mean squared forecast errors (RMSEs) between a given model and the Minnesota VAR while the lower part of the table shows differences in average log predictive likelihoods (LPLs) to the Minnesota VAR. The numbers in the BVAR columns include the actual RMSEs and LPLs of the Minnesota VAR.
\end{minipage}
\end{table}


\subsection{A deeper examination of forecast performance of subspace VAR methods}
To examine more deeply the properties of our subVAR prior, in this sub-section we provide plots over time of predictive Bayes factors against the Minnesota prior VAR. Moreover, we investigate  how the estimates of the prior hyperparameters  $q$, $\omega$ and $\vartheta$ evolve over the hold-out period.

\begin{figure}
\centering
\begin{minipage}[t]{\textwidth}
(a) FEDFUNDS
\end{minipage}
\includegraphics[scale=0.6]{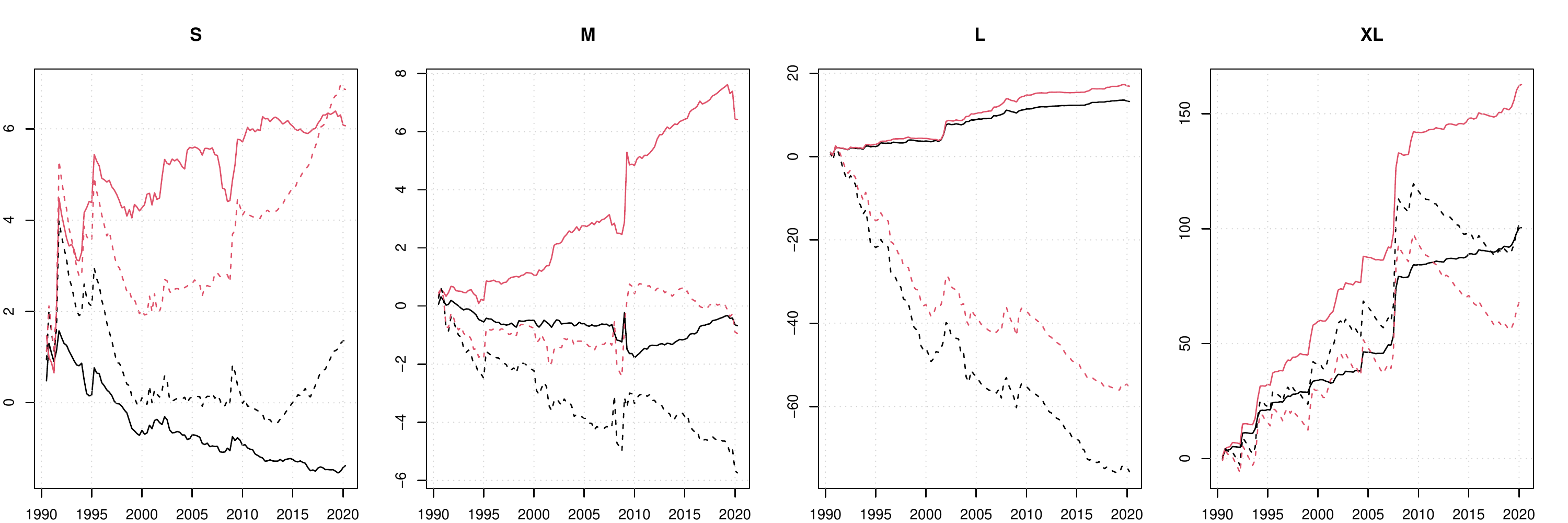}
\begin{minipage}[t]{\textwidth}
(b) CPIAUCSL
\end{minipage}
\includegraphics[scale=0.6]{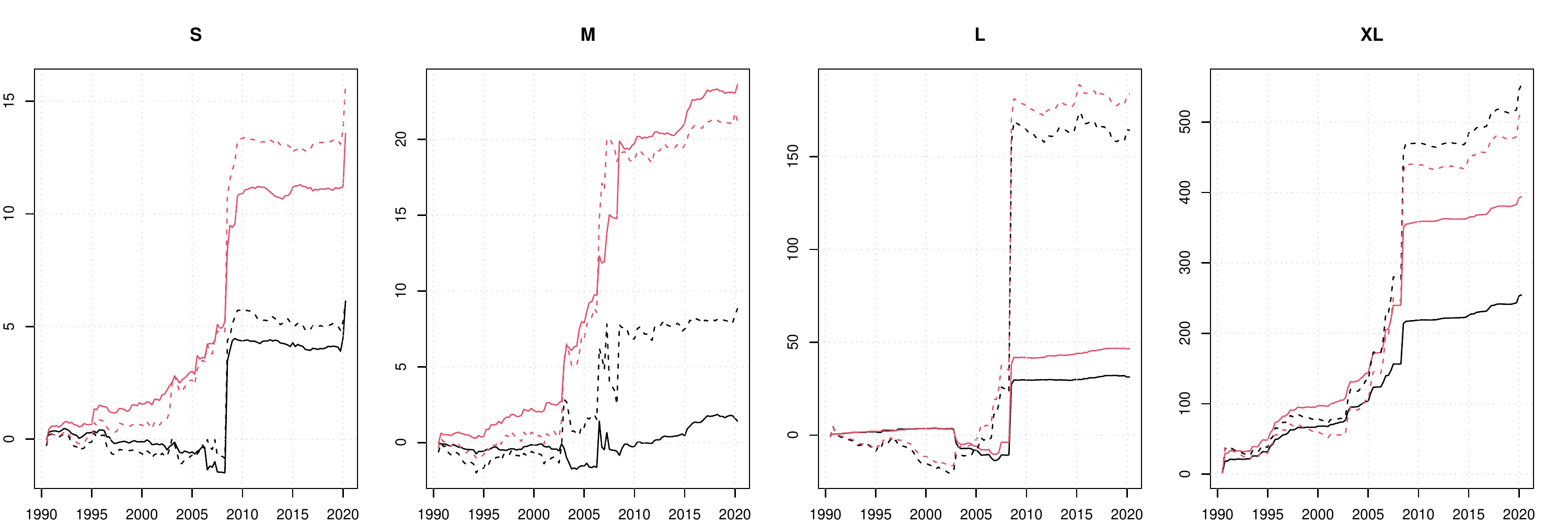}
\begin{minipage}[t]{\textwidth}
(c) GDPC1
\end{minipage}
\includegraphics[scale=0.6]{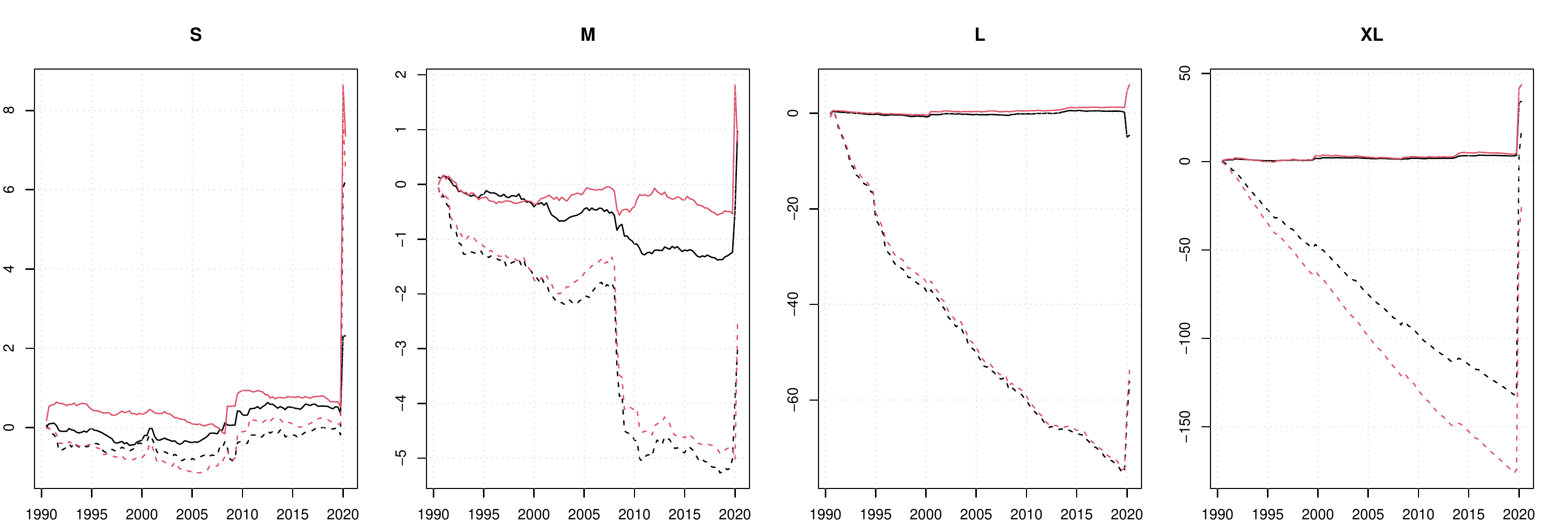}
\begin{minipage}[t]{\textwidth}
\footnotesize \textbf{Notes}: Solid lines refer to the model which includes both subspace and Minnesota shrinkage whereas the dashed lines show the variable-specific log-predictive Bayes factor (to the Minnesota prior VAR) of the subspace shrinkage VAR without the Minnesota prior. Black lines refer to models with an uninformative prior on $\omega$. Red lines denote models with an informative prior on $\omega$.
\end{minipage}
\caption{Evolution of the log predictive Bayes factor between subVAR and the Minnesota VAR across focus variables when the marginal likelihood is used to select $q, \vartheta$ and $\omega$}
\label{lpred_marg}
\end{figure}

\begin{figure}
\centering
\begin{minipage}[t]{\textwidth}
(a) FEDFUNDS
\end{minipage}
\includegraphics[scale=0.6]{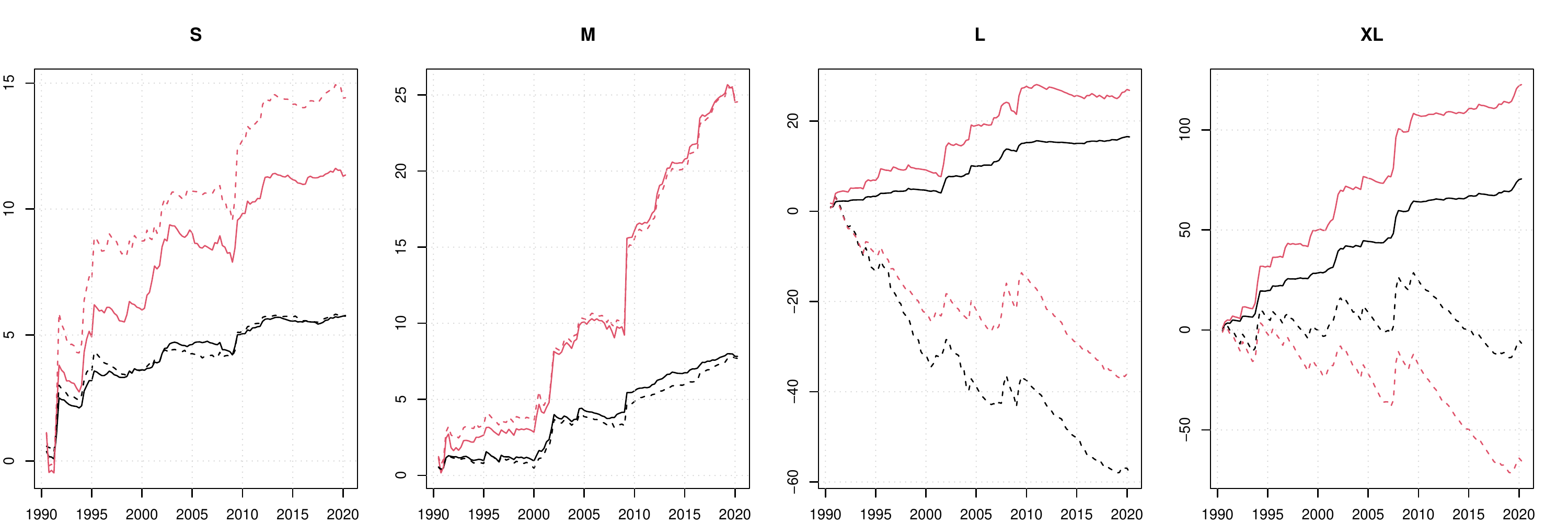}
\begin{minipage}[t]{\textwidth}
(b) CPIAUCSL
\end{minipage}
\includegraphics[scale=0.6]{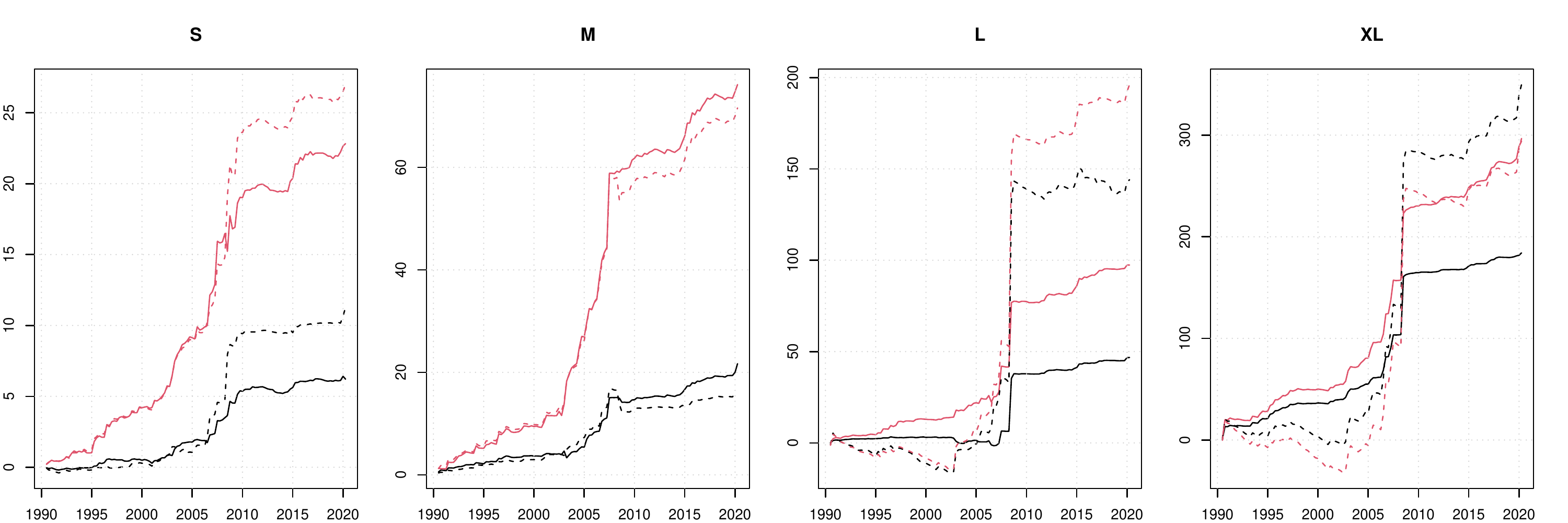}
\begin{minipage}[t]{\textwidth}
(c) GDPC1
\end{minipage}
\includegraphics[scale=0.6]{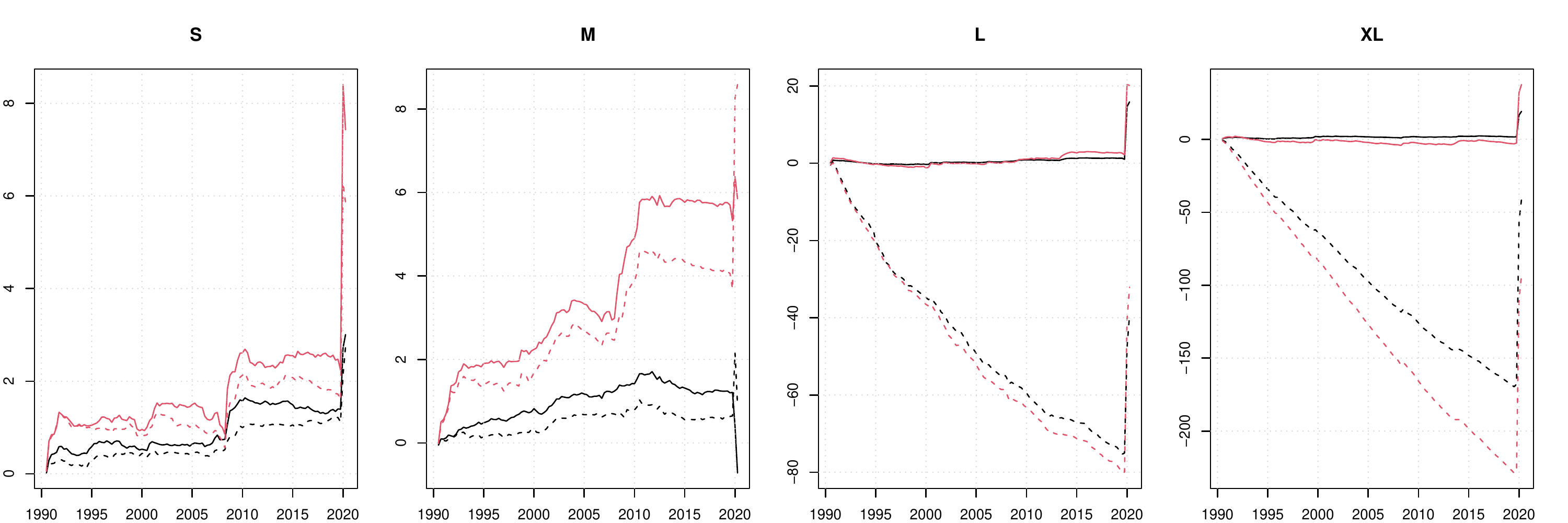}
\begin{minipage}[t]{\textwidth}
\footnotesize \textbf{Notes}: Solid lines refer to the model which includes both subspace and Minnesota shrinkage whereas the dashed lines show the variable-specific log-predictive Bayes factor (to the Minnesota prior VAR) of the subspace shrinkage VAR without the Minnesota prior. Black lines refer to models with an uninformative prior on $\omega$. Red lines denote models with an informative prior on $\omega$.
\end{minipage}
\caption{Evolution of the log predictive Bayes factor between subVAR and the Minnesota VAR across focus variables when the BIC over the three focus variables is used to select $q, \vartheta$ and $\omega$}
\label{lpred_BIC}
\end{figure}

Figures \ref{lpred_marg} and \ref{lpred_BIC} plot the log predictive Bayes factors for the three variables being forecast for the four subspace VAR priors. Figure \ref{lpred_marg} uses the marginal likelihood for all the variables in the model to estimate the prior hyperparameters and Figure \ref{lpred_BIC} uses the BIC for the three variables of interest. The overall best performance of the priors which combine subspace shrinkage with the Minnesota prior can be seen in both figures. An examination of the main exception to this pattern is informative. This occurs for inflation forecasts where the combination of the non-informative prior VAR with the subspace shrinkage prior forecasts well. But this result holds only for the one-quarter-ahead horizon. The iterated one-year-ahead forecasts are very poor (see Table \ref{tab: forecast}).

Another pattern worth noting is that substantial changes tend to occur during either the financial crisis (around 2009) or the pandemic (2020). The tendency at these times is for the subVAR-Flat models to do better. Results for GDP growth from larger VARs are particularly striking. The forecast performance of these models was extremely poor up until the pandemic when the subVAR-Flat models almost caught up to the subVAR-Minn models. The stronger prior information in the latter is a great benefit in normal times, but in the pandemic this makes it less able to adjust to the extreme observations which arise. This is because the corresponding predictive density is narrow which helps in tranquil periods while in turbulent times (such as during the pandemic) the variance is too low, rendering outliers less likely under the posterior predictive distribution.

The role of the prior on $\omega$ can to be seen to be relatively unimportant in most cases. Although it is interesting to note that in smaller models it can be beneficial to use the informative prior for $\omega$ (see, in particular, the forecasts of inflation for the small and medium VARs).

\begin{figure}
\begin{subfigure}[l]{.5\textwidth}
\includegraphics[scale=0.4, page=1]{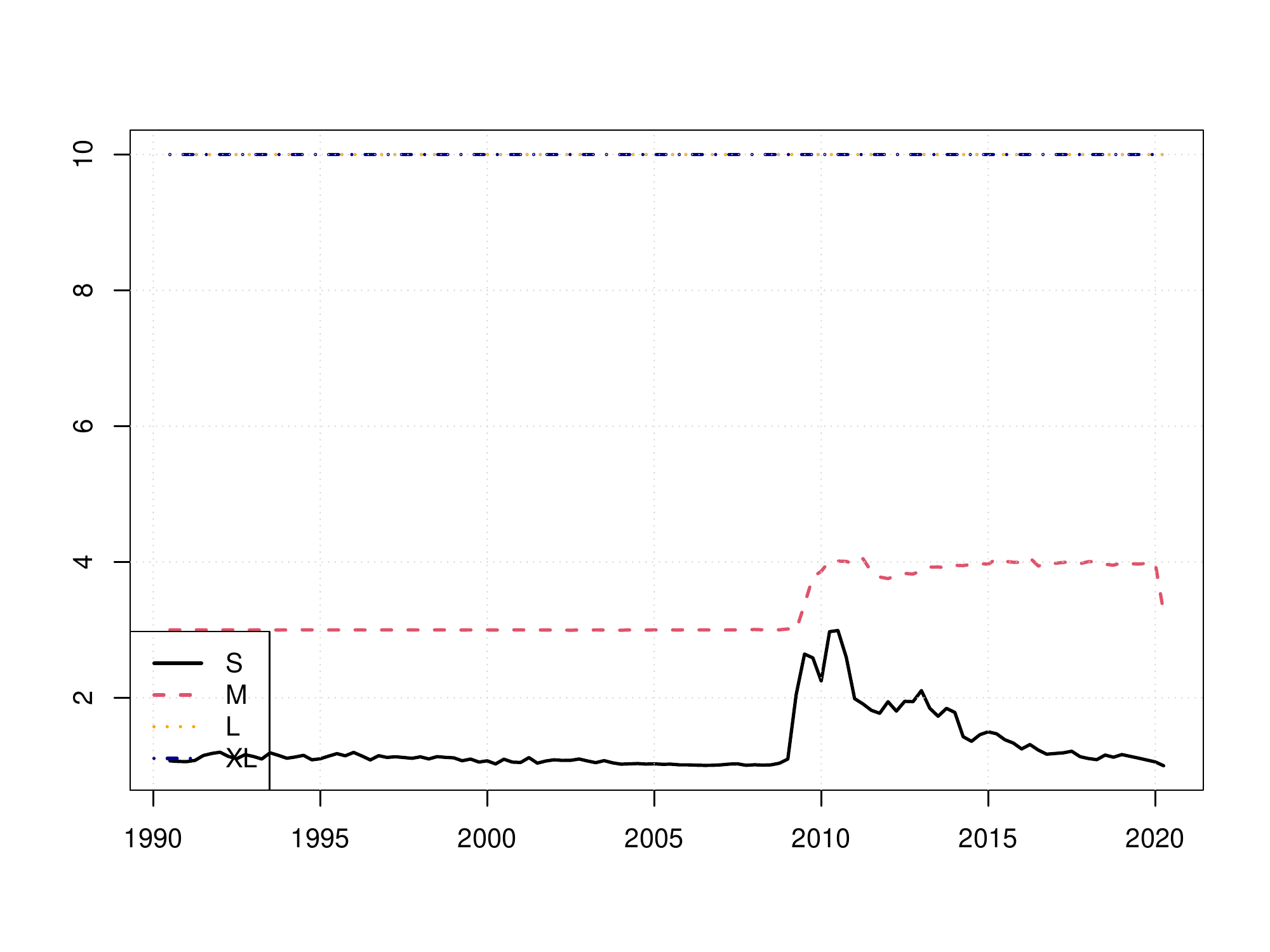} \subcaption{SubVAR-Minn (flat prior on $\omega$)}
\end{subfigure}
\begin{subfigure}[l]{.5\textwidth}
\includegraphics[scale=0.4, page=2]{Results_ML/q_ot.pdf} \subcaption{SubVAR-Minn (informative prior on $\omega$)}
\end{subfigure}\\
\begin{subfigure}[l]{.5\textwidth}
\includegraphics[scale=0.4, page=3]{Results_ML/q_ot.pdf} \subcaption{SubVAR-Flat (flat prior on $\omega$)}
\end{subfigure}
\begin{subfigure}[l]{.5\textwidth}
\includegraphics[scale=0.4, page=4]{Results_ML/q_ot.pdf} \subcaption{SubVAR-Flat (informative prior on $\omega$)}
\end{subfigure}
\caption{Evolution of the posterior mean of $q$ over the hold-out period when the marginal likelihood is used to select $q, \vartheta$ and $\omega$}
\label{q_marg}
\end{figure}

\begin{figure}
\begin{subfigure}[l]{.5\textwidth}
\includegraphics[scale=0.4, page=1]{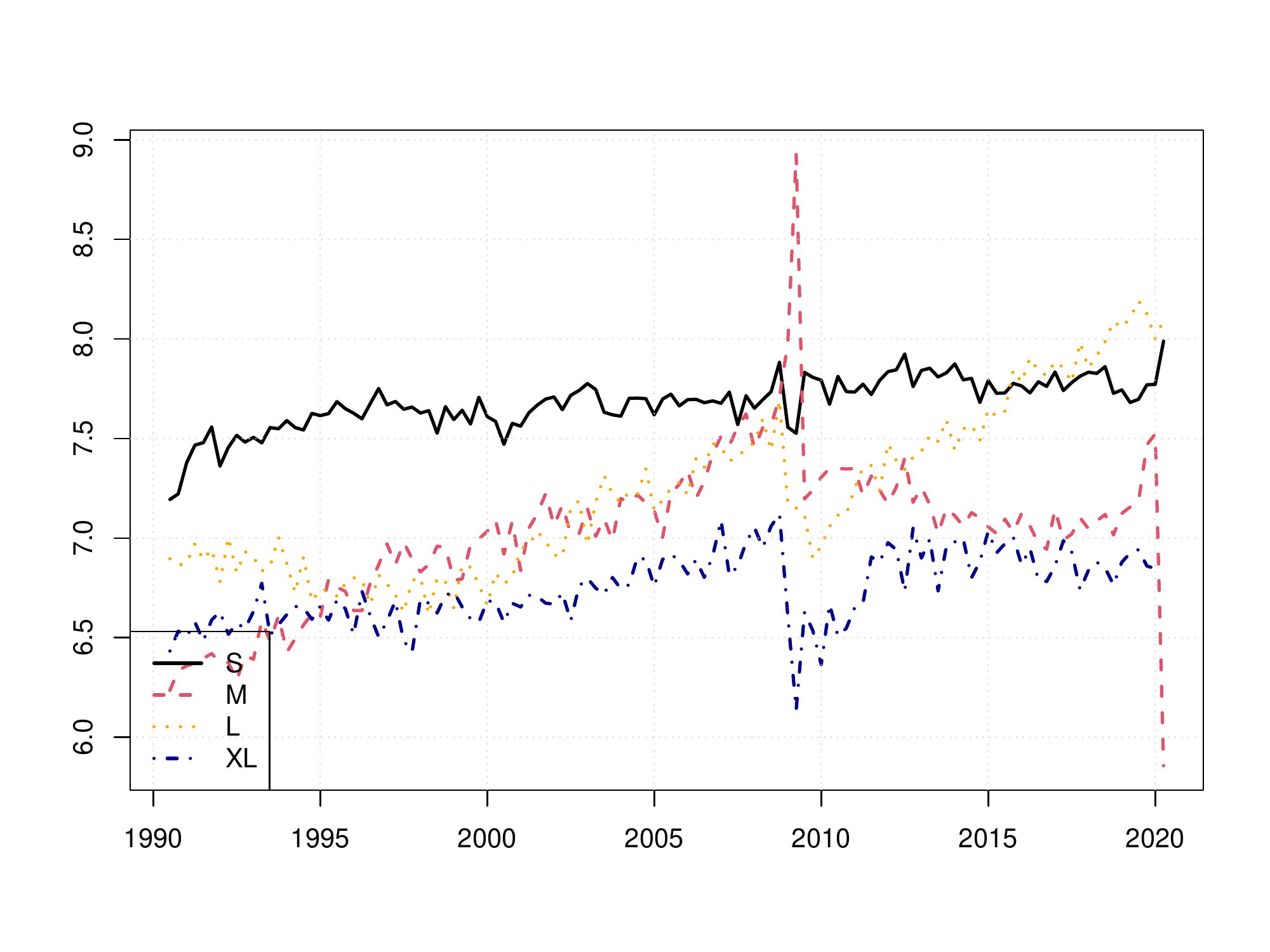} \subcaption{SubVAR-Minn (flat prior on $\omega$)}
\end{subfigure}
\begin{subfigure}[l]{.5\textwidth}
\includegraphics[scale=0.4, page=2]{Results_BIC/q_ot.pdf} \subcaption{SubVAR-Minn (informative prior on $\omega$)}
\end{subfigure}\\
\begin{subfigure}[l]{.5\textwidth}
\includegraphics[scale=0.4, page=3]{Results_BIC/q_ot.pdf} \subcaption{SubVAR-Flat (flat prior on $\omega$)}
\end{subfigure}
\begin{subfigure}[l]{.5\textwidth}
\includegraphics[scale=0.4, page=4]{Results_BIC/q_ot.pdf} \subcaption{SubVAR-Flat (informative prior on $\omega$)}
\end{subfigure}
\caption{Evolution of the posterior mean of $q$ over the hold-out period when the BIC over the three focus variables is used to select $q, \vartheta$ and $\omega$}
\label{q_BIC}
\end{figure}

Figures \ref{q_marg} and \ref{q_BIC} plot the posterior mean of $q$ over time for the marginal likelihood-based and BIC-based methods, respectively. The figures illustrate some interesting differences between these two methods. In particular, use of the BIC allows for more time variation in the parameter estimates for the XL model suggesting it allows for quicker adjustment to new information. Consider the best-performing approach which is the subVAR-Minn prior with an informative prior on $\omega$. For this case, using the marginal likelihood leads to a choice of $10$ factors for all time periods for the XL model. But using the BIC, there is more variation over time. For the XL model in particular the number of factors increases gradually from $7$ to $9$ before quickly collapsing down to a posterior mean near $6$ when the financial crisis hits. In general, for the XL model the marginal likelihood is consistently leading to large estimates for $q$ which vary little over time. This pattern does not recur for the lower dimensional models where the marginal likelihood-based estimates of $q$ tend to be lower. For instance, in the smallest model the sub-VAR flat model chooses $q=1$ for all periods, which contrasts with much larger BIC-based estimates. Marginal likelihood calculation in large VARs can be unstable and sensitive to the prior and our results suggest that in larger models at least it may be safer to use BIC-based estimates.

\begin{figure}
\begin{subfigure}[l]{.5\textwidth}
\includegraphics[scale=0.4, page=1]{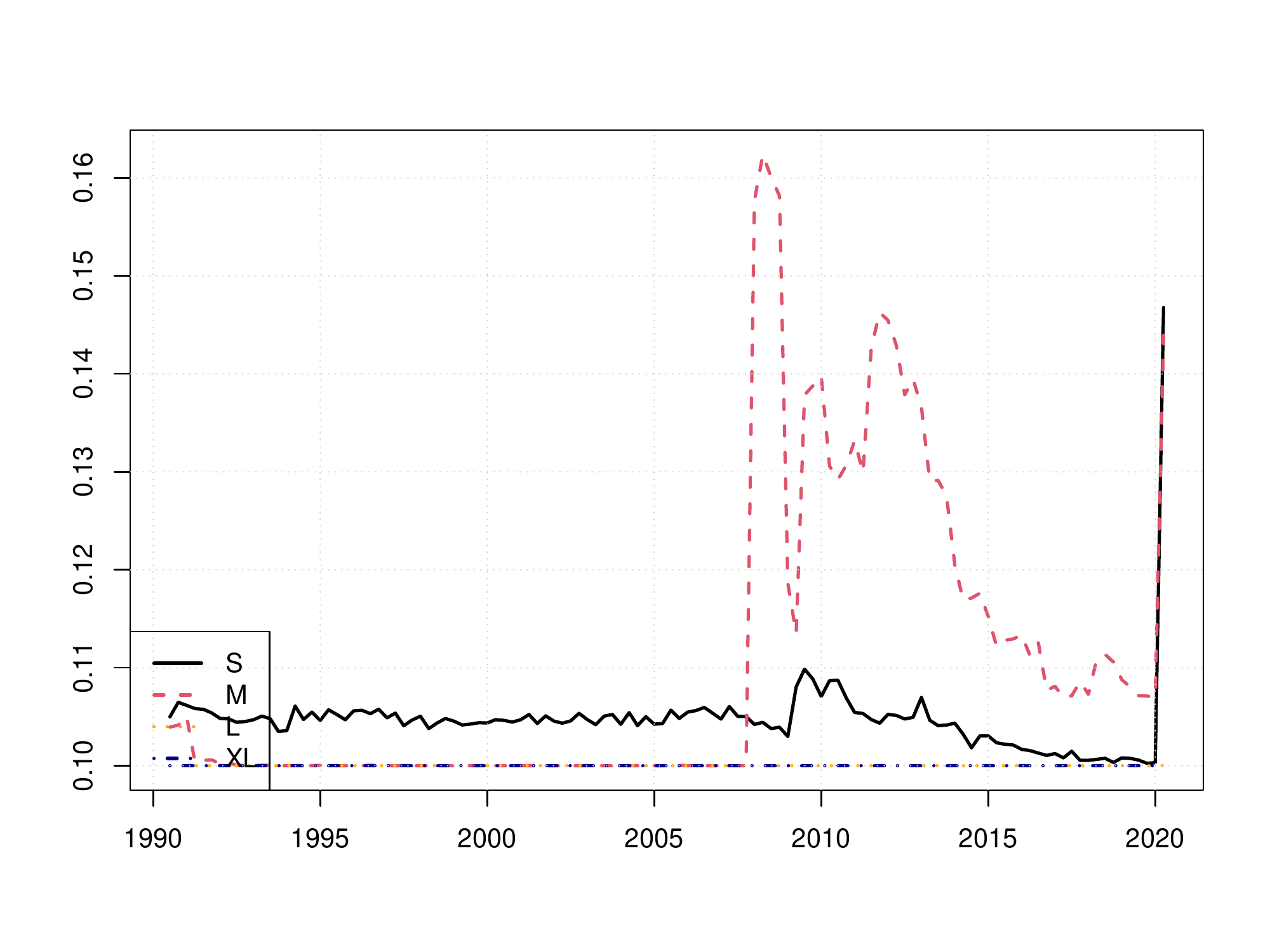} \subcaption{SubVAR-Minn (flat prior on $\omega$)}
\end{subfigure}
\begin{subfigure}[l]{.5\textwidth}
\includegraphics[scale=0.4, page=2]{Results_ML/omega_ot.pdf} \subcaption{SubVAR-Minn (informative prior on $\omega$)}
\end{subfigure}\\
\begin{subfigure}[l]{.5\textwidth}
\includegraphics[scale=0.4, page=3]{Results_ML/omega_ot.pdf} \subcaption{SubVAR-Flat (flat prior on $\omega$)}
\end{subfigure}
\begin{subfigure}[l]{.5\textwidth}
\includegraphics[scale=0.4, page=4]{Results_ML/omega_ot.pdf} \subcaption{SubVAR-Flat (informative prior on $\omega$)}
\end{subfigure}
\caption{Evolution of the posterior mean of $\omega$ over the hold-out period when the marginal likelihood is used to select $q, \vartheta$ and $\omega$}
\label{omega_marg}
\end{figure}

\begin{figure}
\begin{subfigure}[l]{.5\textwidth}
\includegraphics[scale=0.4, page=1]{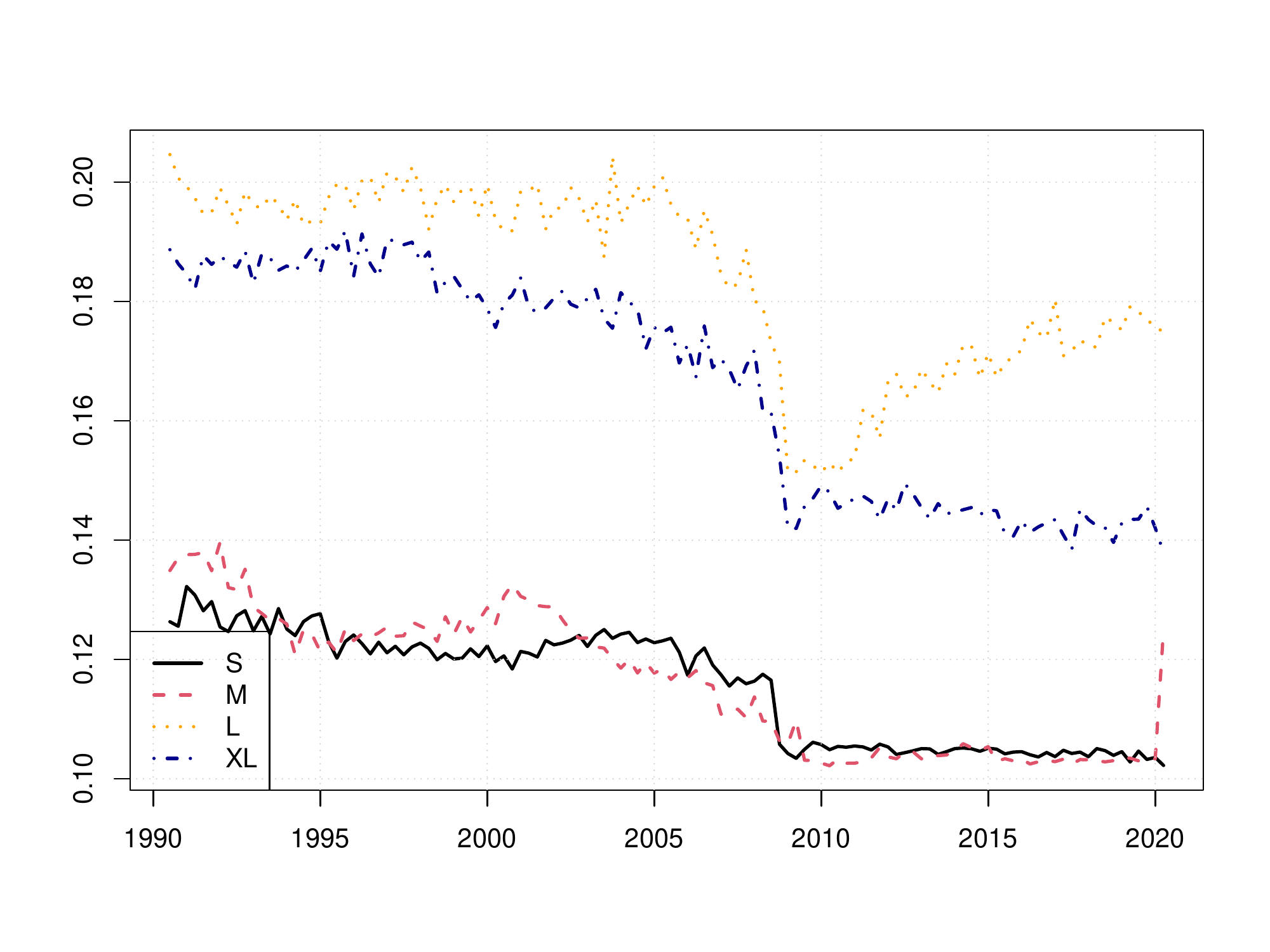} \subcaption{SubVAR-Minn (flat prior on $\omega$)}
\end{subfigure}
\begin{subfigure}[l]{.5\textwidth}
\includegraphics[scale=0.4, page=2]{Results_BIC/omega_ot.pdf} \subcaption{SubVAR-Minn (informative prior on $\omega$)}
\end{subfigure}\\
\begin{subfigure}[l]{.5\textwidth}
\includegraphics[scale=0.4, page=3]{Results_BIC/omega_ot.pdf} \subcaption{SubVAR-Flat (flat prior on $\omega$)}
\end{subfigure}
\begin{subfigure}[l]{.5\textwidth}
\includegraphics[scale=0.4, page=4]{Results_BIC/omega_ot.pdf} \subcaption{SubVAR-Flat (informative prior on $\omega$)}
\end{subfigure}
\caption{Evolution of the posterior mean of $\omega$ over the hold-out period when the BIC over the three focus variables is used to select $q, \vartheta$ and $\omega$}
\label{omega_BIC}
\end{figure}

Figures \ref{omega_marg} and \ref{omega_BIC} present evidence on the estimation of $\omega$. For the XL and L models, we are finding striking differences between the BIC and marginal-likelihood based estimates. Note that for subVAR-Minn model with the informative prior on $\omega$ we are finding the posterior mean of $\omega$ to be approximately $0.6$ when estimated using BIC whereas the marginal likelihood based estimates are much lower at approximately $0.25$/$0.35$ for the XL/L models. Hence, the former model is shrinking much more closely to the factor model than the latter. 

We can also see the role that the prior for $\omega$ has in that estimates using the non-informative prior for $\omega$ tend to be substantially lower than those produced using the informative prior. In fact, with rare exceptions, using the non-informative prior never leads to estimates of $\omega$ above $0.2$. At least in this data set, it is necessary to use an informative prior for $\omega$ to achieve substantial shrinkage towards the factor model. It is interesting to note that, when we do so, we are consistently finding $\omega$ to be in the region $[0.25,0.60]$ being far from the region where one would feel confident selecting either the Minnesota prior VAR ($\omega=0$) or the factor model ($\omega=1$) thus indicating again the potential benefits of our approach which combines the two.    

\begin{figure}
\begin{subfigure}[l]{.5\textwidth}
\includegraphics[scale=0.4, page=1]{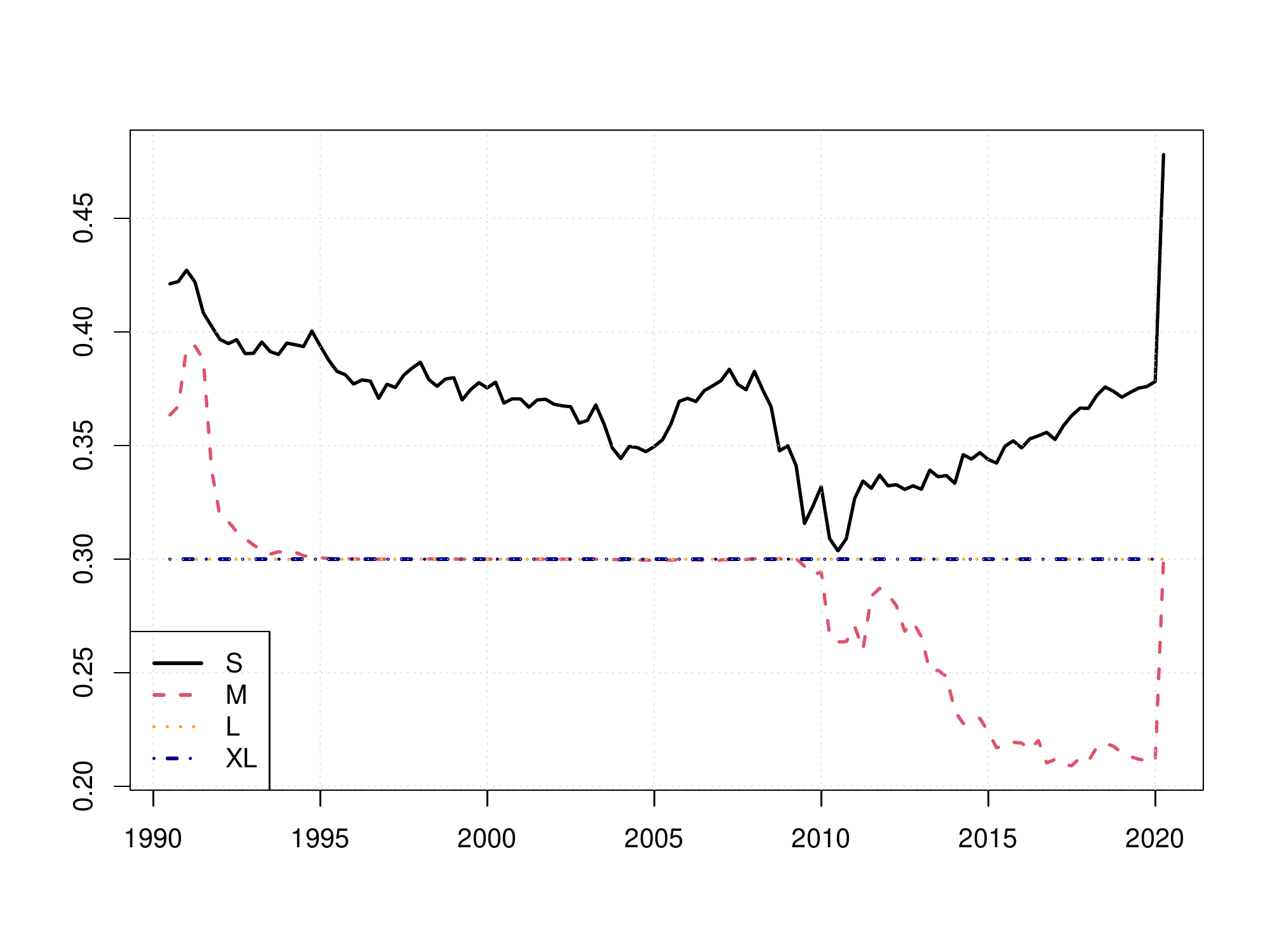} \subcaption{SubVAR-Minn (flat prior on $\omega$, ML)}
\end{subfigure}
\begin{subfigure}[l]{.5\textwidth}
\includegraphics[scale=0.4, page=1]{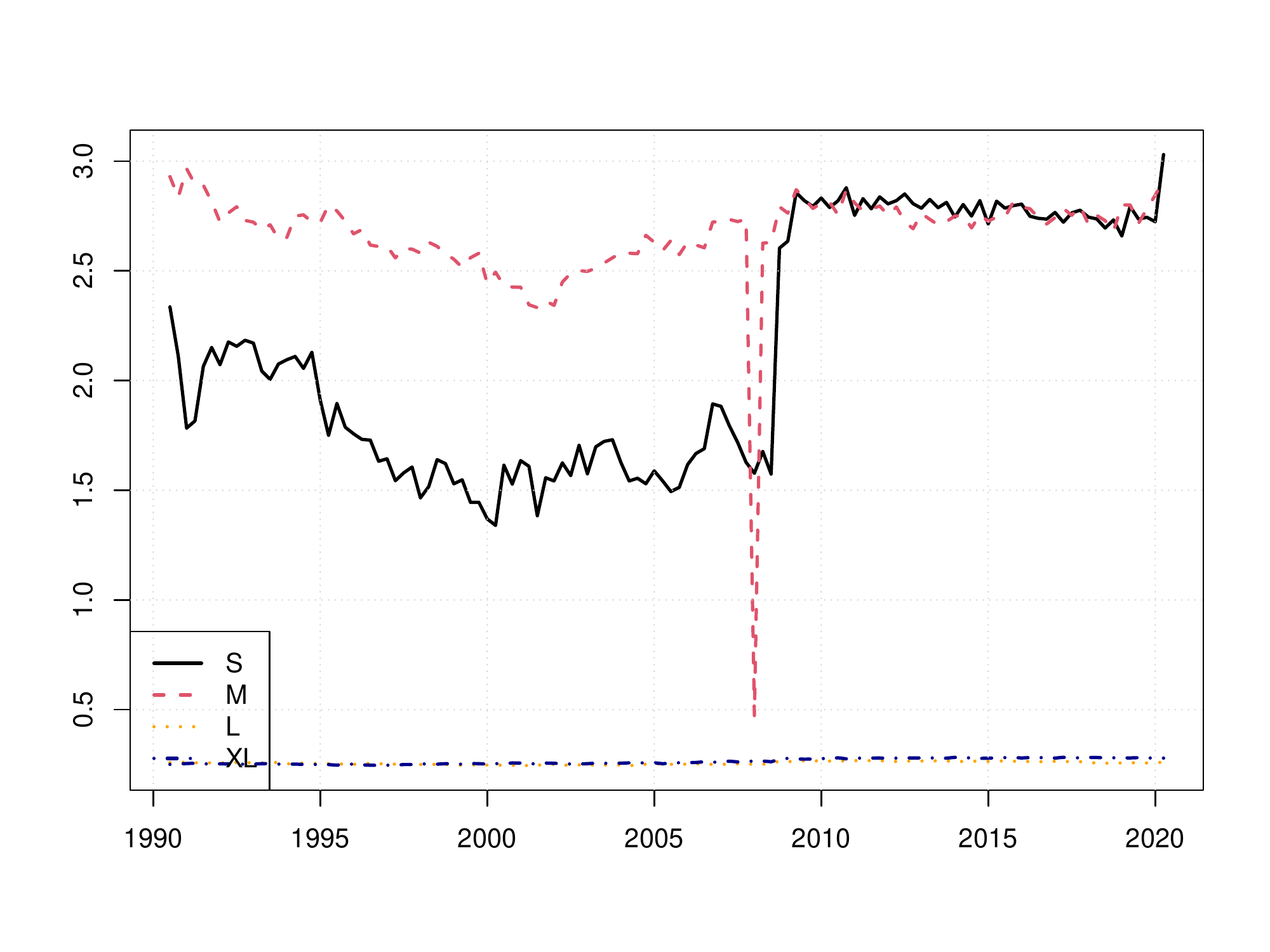} \subcaption{SubVAR-Minn (flat prior on $\omega$, BIC)}
\end{subfigure}
\begin{subfigure}[l]{.5\textwidth}
\includegraphics[scale=0.4, page=2]{Results_ML/theta_ot.pdf} \subcaption{SubVAR-Minn (informative prior on $\omega$, ML)}
\end{subfigure}
\begin{subfigure}[l]{.5\textwidth}
\includegraphics[scale=0.4, page=2]{Results_BIC/theta_ot.pdf} \subcaption{SubVAR-Minn (informative prior on $\omega$, BIC)}
\end{subfigure}
\caption{Evolution of the posterior mean of $\vartheta$ over the hold-out period}
\label{vartheta}
\end{figure}

Finally we turn to main shrinkage parameter of the Minnesota prior, $\vartheta$. This hyperparameter only appears in the approaches involving the Minnesota prior. Note that smaller values of $\vartheta$ imply stronger shrinkage. Posterior means are plotted in Figure \ref{vartheta}.

The most striking pattern here is that using the marginal likelihood leads to much lower estimates of this hyperparameter than using the BIC, especially for the small and medium models. In general, and consistent with \cite{GiannoneLenzaPrimiceri2015RESTAT}, we find that larger models generally feature smaller values of $\vartheta$ (and thus more shrinkage). If we combine this with the fact that the marginal likelihood-based estimates of $\omega$ are lower for these models we have the interesting finding that it is choosing to put more weight on a Minnesota prior VAR with more shrinkage. In contrast, the BIC based weights are closer to be a combination of a factor model with a Minnesota prior that is implemented rather loosely.   

Another interesting finding is that $\vartheta$ tends to sharply increase  during the pandemic. This is especially pronounced for small and medium-sized models. Since the variance of the predictive distribution is positively related to $\vartheta$, larger values of $\vartheta$ are (all else being equal) accompanied by wider predictive intervals. This explains why some of the models improve appreciable against the benchmark  in 2020.

\section{Further Discussion}
\label{sec_extensions}

In this paper, we have worked with two popular models (i.e., the conjugate version of the Minnesota prior VAR with a single shrinkage hyperparameter and the factor model) both of which are homoskedastic. We did this to draw out all the theoretical insights in a clear and simple way and because, in many empirical contexts, simple approaches such as these have been found to work well (see, e.g., \cite{BanburaGiannoneReichlin:2010:jae:large:bvar} and \cite{ccm2015}). Furthermore, computation is vastly simplified since analytical results are available and we can avoid the use of MCMC methods. 

However, many recent Bayesian VAR papers have used richer econometric structures. These can be classified in two main categories: other priors and other forms for the error covariance matrix. Here we discuss using the subVAR prior in the context of such extensions.

One restrictive feature of our Minnesota prior is that it involves a single shrinkage parameter. Allowing for each equation to have its own shrinkage parameter could be a useful extension of our approach. This raises two issues: i) the resulting priors would no longer be conjugate and ii) the number of prior hyperparameters would become larger leading to a higher dimensional grid of values to evaluate in the Monte Carlo step. These issues could partly be surmounted by decomposing $\bm \Sigma = \bm A_0^{-1}(\bm A_0^{-1})'$ and writing the VAR in structural form (i.e. with 
$\bm A_0 \bm Y_t$ on the left hand side and the error covariance matrix being diagonal). Conditional on $\bm A_0$, the subspace shrinkage prior could be applied in an equation-specific manner and estimation could proceed one equation at a time (which would help reduce the computational burden, although the computational cost of having $M$ different sets of prior hyperparameters would still be large). $\bm A_0$ could be drawn using MCMC methods. In essence, it would be straightforward to develop an MCMC algorithm for drawing $\bm A_0$, $\omega_j$, $q_j$ and $\vartheta_j$ for $j=1,..,M$. Being an MCMC algorithm it would be inherently much more computationally demanding than the methods developed in this paper, but could be useful at least in medium-sized VARs. 

There have also been many global-local shrinkage priors (e.g., the Horseshoe and Lasso priors) which are conditionally Gaussian (i.e., conditional on some new parameters in the prior they are Gaussian). Estimation proceeds by adding blocks to the MCMC algorithm for drawing these new parameters. Since our Minnesota prior is Gaussian, it is trivial to replace it with any conditionally Gaussian prior. The theory developed in this paper would hold, conditional on the new parameters. Estimation would proceed through an MCMC algorithm which drew these new parameters and then conditional on each draw exploited the subVAR methods developed in this paper. 

In a similar fashion, the assumption of homoskedasticity could be relaxed to allow for stochastic volatility. This would lead to an MCMC algorithm which involved drawing the volatilities and, conditional on each draw,  the results for the subspace VAR prior developed in this paper could be used. Several forms for stochastic  volatility in VARs have been proposed in the literature, see for instance \cite{CarrieroClarkMarcellino:2019:joe:VAR:triang} for a particularly popular form, and the general strategy outlined here would work with any of them. 

In sum, many extensions of the conjugate subVAR approach developed in this paper are possible. However, they would require the use of MCMC methods. Provided the likelihood and prior remain Gaussian conditional on some new parameters, the theory derived in Section 2 would hold, conditional on these new parameters.  

Finally, it is worth noting that the error covariance structure proposed in \cite{Chan2020} maintains many of the benefits of the conjugate form. It assumes the VAR error covariance matrix is $\bm \Sigma \otimes \bm \Omega$ where $\bm \Omega$ can be any positive definite matrix. This nests many possible specifications, including a common stochastic volatility model, moving average errors and non-Gaussian errors. This Kronecker structure in the likelihood matches up with the Kronecker structure in the conjugate prior leading  to derivations which are similar to those in Section 2 of this paper. Roughly speaking, whereas the derivations in Sub-section \ref{sec: sub.vars} show how the subspace prior leads to a posterior mean which is a combination of the OLS estimate of the VAR with a PC regression, using the model of \cite{Chan2020} leads to a combination of a GLS estimate with a PC regression. \cite{Chan2020} develops a computationally efficient MCMC algorithm for models with this error covariance structure. 

\section{Conclusions}
Macroeconomic researchers with large data sets have traditionally been forced to make a choice between a large VAR or a dynamic factor model. In this paper, we have shown how to combine the two. We have developed a subspace prior for the VAR which shrinks towards a dynamic factor model. A parameter, $\omega$ controls the degree of shrinkage and we have developed methods for estimating it from the data. Thus, we have developed a Bayesian methodology for averaging a large VAR with a factor model or choosing between them. 

We illustrate our approach using synthetic and real data. In simulations, we show that our approach accurately detects the number of factors if the true DGP suggests relatively few factors (irrespective of the model size). If the DGP features a large number of factors, our approach slightly underestimates the true number of factors.  In a forecasting exercise involving a large number of macroeconomic variables, we demonstrate the benefits of combining the two model classes using our subspace VAR approach. Using subspace shrinkage in combination with a Minnesota prior often yields more precise forecasts than the ones obtained from either the factor model or the VAR.




\clearpage\small{\setstretch{0.85}
\addcontentsline{toc}{section}{References}
\bibliographystyle{frbcle.bst}
\bibliography{lit}}\normalsize\clearpage

@article{huber2019adaptive,
  title={Adaptive shrinkage in Bayesian vector autoregressive models},
  author={Huber, Florian and Feldkircher, Martin},
  journal={Journal of Business \& Economic Statistics},
  volume={37},
  number={1},
  pages={27--39},
  year={2019},
  publisher={Taylor \& Francis}
}

@article{korobilis2019adaptive,
  title={Adaptive hierarchical priors for high-dimensional vector autoregressions},
  author={Korobilis, Dimitris and Pettenuzzo, Davide},
  journal={Journal of Econometrics},
  volume={212},
  number={1},
  pages={241--271},
  year={2019},
  publisher={Elsevier}
}

@article{sims1998bayesian,
  title={Bayesian methods for dynamic multivariate models},
  author={Sims, Christopher A and Zha, Tao},
  journal={International Economic Review},
  pages={949--968},
  year={1998},
  publisher={JSTOR}
}

@article{kadiyala1997numerical,
  title={Numerical methods for estimation and inference in Bayesian VAR-models},
  author={Kadiyala, K Rao and Karlsson, Sune},
  journal={Journal of Applied Econometrics},
  volume={12},
  number={2},
  pages={99--132},
  year={1997},
  publisher={Wiley Online Library}
}

@techreport{mccracken2020fred,
  title={FRED-QD: A quarterly database for macroeconomic research},
  author={McCracken, Michael and Ng, Serena},
  year={2020},
  institution={National Bureau of Economic Research}
}

@article{hauzenberger2021combining,
  title={Combining shrinkage and sparsity in conjugate vector autoregressive models},
  author={Hauzenberger, Niko and Huber, Florian and Onorante, Luca},
  journal={Journal of Applied Econometrics},
  volume={36},
  number={3},
  pages={304--327},
  year={2021},
  publisher={Wiley Online Library}
}

@article{giannone2019priors,
  title={Priors for the long run},
  author={Giannone, Domenico and Lenza, Michele and Primiceri, Giorgio E},
  journal={Journal of the American Statistical Association},
  volume={114},
  number={526},
  pages={565--580},
  year={2019},
  publisher={Taylor \& Francis}
}

@article{subspace,
    author = {Minsuk Shin and Anirban Bhattacharya and Valen E. Johnson},
    title = {Functional horseshoe priors for subspace shrinkage},
    journal = {Journal of the American Statistical Association},
    year = {2020},
    volume = {115},
    pages = {1784-1797}
}

@article{Minnesota1,
    author = {Tom Doan and Robert Litterman and Chris Sims},
    title = {Forecasting and conditional projections using realistic prior distributions},
    journal = {Econometric Reviews},
    year = {1984},
    volume = {3},
    pages = {1-144}
}

@article{Minnesota2,
    author = {Robert Litterman},
    title = {Forecasting with Bayesian Vector Autoregressions - Fuve years of experience},
    journal = {Journal of Business and Economic Statistics},
    year = {1986},
    volume = {4},
    pages = {25-38}
}

@article{LWfactor,
    author = {Hedibert Lopes and Mike West},
    title = {Bayesian model assessment in factor analysis},
    journal = {Statistica Sinica},
    year = {2004},
    volume = {4},
    pages = {41-67}
}

@article{ccm2015,
    author = {Andrea Carriero and Todd E. Clark and Massimiliano Marcellino},
    title = {Bayesian VARs: Specification choices and  forecast accuracy},
    journal = {Journal of Applied Econometrics},
    year = {2015},
    volume = {30},
    pages = {46-73}
}

@article{Lopes_FS,
  title={Sparse Bayesian Factor Analysis when the Number of Factors is Unknown},
  author={Fr{\"u}hwirth-Schnatter, Sylvia and Lopes, Hedibert},
  journal={arXiv},
    volume={https://arxiv.org/abs/1804.04231},
  year={2018},
  
}

@article{oman1982,
  title={Shrinking towards subspaces in multiple linear regression},
  author={Oman, Samuel},
  journal={Technometrics},
  volume={24},
  pages={307--311},
  year={1982}
}

@article{BD2011,
  title={Sparse Bayesian infinite factor models},
  author={Bhattacharya, A. and Dunson, D. },
  journal={Biometrika},
  volume={98},
  pages={291--306},
  year={2011},
 }

@article{stock2002macroeconomic,
  title={Macroeconomic forecasting using diffusion indexes},
  author={Stock, James H and Watson, Mark W},
  journal={Journal of Business \& Economic Statistics},
  volume={20},
  number={2},
  pages={147--162},
  year={2002},
  doi = {},
  publisher={Taylor \& Francis}
}

@incollection{geweke1977,
  title={The dynamic factor analysis of economic time series},
  author={Geweke, John},
  booktitle={Latent Variables in Socio-economic Models},
  year={1977},
  publisher={North-Holland, Amsterdam}
}

@book{BEM2,
  title={Bayesian Econometric Methods},
  author={Chan, Joshua and Koop, Gary and Tobias, Justin and Poirier, Dale},
  year={2019},
  publisher={Cambridge University Press}
}

@article{BBE,
 title ={Measuring the effects of monetary policy: A Factor Augmented Vector Autoregressive ({FAVAR}) approach},
 author={Bernanke,Ben and Boivin, Jean and Eliasz, Piotr},
 journal={Quarterly Journal of Economics},
 volume={120},
  pages={387--422},
  year={2005}
}

@Article{GiannoneLenzaPrimiceri2015RESTAT,
  author={Domenico Giannone and Michele Lenza and Giorgio E. Primiceri},
  title={Prior Selection for Vector Autoregressions},
  journal={The Review of Economics and Statistics},
  year=2015,
  volume={97},
  number={2},
  pages={436-451},
  month={May},
  doi={10.1162/rest_a_00483},
  url={https://ideas.repec.org/a/tpr/restat/v97y2015i2p436-451.html}
}

@Article{BanburaGiannoneReichlin:2010:jae:large:bvar,
  author={Marta Banbura and Domenico Giannone and Lucrezia Reichlin},
  title={Large {B}ayesian vector auto regressions},
  journal={Journal of Applied Econometrics},
  year=2010,
  volume={25},
  number={1},
  pages={71-92},
  doi={10.1002/jae.1137},
  url={https://ideas.repec.org/a/jae/japmet/v25y2010i1p71-92.html}
}

@Article{KoopKorobilis2019,
  author={Gary Koop and Dimitris Korobilis},
  title={Forecasting with high dimensional panel VARs},
  journal={Oxford Bulletin of Economics and Statistics},
  year={2019},
  volume={81},
  pages={937-959},
  
}

@Article{Jarocinski2017,
  author={Marek Jarocinski and Bartosz Mackowiak},
  title={ Granger-causal-priority and choice of variables in vector autoregressions},
  journal={Review of Economics and Statistics},
  year={2017},
  volume={99},
  pages={319-329},
  
}

@Article{Chan2020,
  author={Joshua Chan},
  title={Large Bayesian {VARs}: A Flexible Kronecker Error Covariance Structure},
  journal={Journal of Businss and Economic Statistics},
  year={2020},
  volume={38},
  pages={68-79},
  
}

@Article{Koop:2013:jae:medium:bvar,
  author={Gary Koop},
  title={Forecasting with Medium and Large {B}ayesian {VARS}},
  journal={Journal of Applied Econometrics},
  year=2013,
  volume={28},
  number={2},
  pages={177-203},
  month=mar,
  doi={10.1002/jae.1270},
  url={https://ideas.repec.org/a/wly/japmet/v28y2013i2p177-203.html}
}

@Article{CarrieroClarkMarcellino:2019:joe:VAR:triang,
  author={Carriero, Andrea and Clark, {Todd E.} and Marcellino, Massimiliano},
  title={Large {B}ayesian vector autoregressions with stochastic volatility and non-conjugate priors},
  journal={Journal of Econometrics},
  year=2019,
  volume=212,
  number=1,
  pages={137-154},
  doi={10.1016/j.jeconom.2019.04.024},
  url={https://ideas.repec.org/a/eee/econom/v212y2019i1p137-154.html}
}

\begin{appendices}
\section{Data Appendix}
\begin{table}[h!]
\caption{Description of the Dataset} \label{tab: dataset}
\scalebox{0.5}{
\begin{tabular}{lllccccc} 
  \hline
  ID & FRED Code & Description & Transformation Codes & S & M & L & XL \\ 
  \hline
1 & GDPC1 & Real Gross Domestic Product & 5 & X & X & X & X \\ 
  2 & PCECC96 & Real Personal Consumption Expenditures & 5 &  & X & X & X \\ 
  3 & PCDGx & Real personal consumption expenditures:  Durable goods  & 5 &  &  &  & X \\ 
  4 & PCESVx & Real Personal Consumption Expenditures:  Services  & 5 &  &  & X & X \\ 
  5 & PCNDx & Real Personal Consumption Expenditures:  Nondurable Goods  & 5 &  &  & X & X \\ 
  6 & GPDIC1 & Real Gross Private Domestic Investment & 5 &  &  & X & X \\ 
  7 & FPIx & Real private fixed investment  & 5 &  & X & X & X \\ 
  8 & Y033RC1Q027SBEAx & Real Gross Private Domestic Investment:  Fixed Investment:  Nonresidential Equipment \hspace{7cm} & 5 &  &  & X & X \\ 
  9 & PNFIx & Real private fixed investment:  Nonresidential  & 5 &  &  & X & X \\ 
  10 & PRFIx & Real private fixed investment:  Residential  & 5 &  &  & X & X \\ 
  11 & A014RE1Q156NBEA & Shares of gross domestic product:  Gross private domestic investment: Change
in private inventories & 1 &  &  & X & X \\ 
  12 & GCEC1 & Real Government Consumption Expenditures and Gross Investment & 5 &  & X & X & X \\ 
  13 & A823RL1Q225SBEA & Real Government Consumption Expenditures and Gross Investment:  Federal & 1 &  &  &  & X \\ 
  14 & FGRECPTx & Real Federal Government Current Receipts  & 5 &  &  &  & X \\ 
  15 & SLCEx & Real government state and local consumption expenditures  & 5 &  &  &  & X \\ 
  16 & EXPGSC1 & Real Exports of Goods and Services & 5 &  &  & X & X \\ 
  17 & IMPGSC1 & Real Imports of Goods and Services & 5 &  &  & X & X \\ 
  18 & DPIC96 & Real Disposable Personal Income & 5 &  &  & X & X \\ 
  19 & OUTNFB & Nonfarm Business Sector:  Real Output & 5 &  &  &  & X \\ 
  20 & OUTBS & Business Sector:  Real Output & 5 &  &  &  & X \\ 
  21 & INDPRO & IP:Total index Industrial Production Index (Index 2012=100) & 5 &  & X & X & X \\ 
  22 & IPFINAL & IP:Final products Industrial Production: Final Products (Market Group) (Index 2012=100) & 5 &  &  & X & X \\ 
  23 & IPCONGD & IP:Consumer goods Industrial Production: Consumer Goods (Index 2012=100) & 5 &  &  & X & X \\ 
  24 & IPMAT & Materials (Index 2012=100) & 5 &  &  &  & X \\ 
  25 & IPDMAT & Durable Materials (Index 2012=100) & 5 &  &  &  & X \\ 
  26 & IPNMAT & Nondurable Materials (Index 2012=100) & 5 &  &  &  & X \\ 
  27 & IPDCONGD & Durable Consumer Goods (Index 2012=100) & 5 &  &  &  & X \\ 
  28 & IPB51110SQ & Durable Goods:  Automotive products (Index 2012=100) & 5 &  &  &  & X \\ 
  29 & IPNCONGD & Nondurable Consumer Goods (Index 2012=100) & 5 &  &  &  & X \\ 
  30 & IPBUSEQ & Business Equipment (Index 2012=100) & 5 &  &  &  & X \\ 
  31 & IPB51220SQ & Consumer energy products (Index 2012=100) & 5 &  &  &  & X \\ 
  32 & CUMFNS & Capacity Utilization:  Manufacturing (SIC) (Percent of Capacity) & 1 &  &  &  & X \\ 
  33 & IPMANSICS & Industrial Production:  Manufacturing (SIC) (Index 2012=100) & 5 &  &  &  & X \\ 
  34 & IPB51222S & Industrial Production:  Residential Utilities (Index 2012=100) & 5 &  &  &  & X \\ 
  35 & IPFUELS & Industrial Production:  Fuels (Index 2012=100) & 5 &  &  &  & X \\ 
  36 & PAYEMS &  Emp:Nonfarm All Employees: Total nonfarm (Thousands of Persons) & 5 &  &  & X & X \\ 
  37 & USPRIV &  All Employees: Total Private Industries (Thousands of Persons) & 5 &  &  &  & X \\ 
  38 & MANEMP &  All Employees: Manufacturing (Thousands of Persons) & 5 &  &  &  & X \\ 
  39 & SRVPRD & All Employees:  Service-Providing Industries (Thousands of Persons) & 5 &  &  &  & X \\ 
  40 & USGOOD & All Employees:  Goods-Producing Industries (Thousands of Persons) & 5 &  &  &  & X \\ 
  41 & DMANEMP & All Employees:  Durable goods (Thousands of Persons) & 5 &  &  &  & X \\ 
  42 & NDMANEMP & All Employees:  Nondurable goods (Thousands of Persons) & 5 &  &  &  & X \\ 
  43 & USCONS & All Employees:  Construction (Thousands of Persons) & 5 &  &  &  & X \\ 
  44 & USEHS & All Employees:  Education \& Health Services (Thousands of Persons) & 5 &  &  &  & X \\ 
  45 & USFIRE & All Employees:  Financial Activities (Thousands of Persons) & 5 &  &  &  & X \\ 
  46 & USINFO & All Employees:  Information Services (Thousands of Persons) & 5 &  &  &  & X \\ 
  47 & USPBS & All Employees:  Professional \& Business Services (Thousands of Persons) & 5 &  &  &  & X \\ 
  48 & USLAH & All Employees:  Leisure \& Hospitality (Thousands of Persons) & 5 &  &  &  & X \\ 
  49 & USSERV & All Employees:  Other Services (Thousands of Persons) & 5 &  &  &  & X \\ 
  50 & USMINE & All Employees:  Mining and logging (Thousands of Persons) & 5 &  &  &  & X \\ 
  51 & USTPU & All Employees:  Trade, Transportation \& Utilities (Thousands of Persons) & 5 &  &  &  & X \\ 
  52 & USGOVT & All Employees:  Government (Thousands of Persons) & 5 &  &  &  & X \\ 
  53 & USTRADE & All Employees:  Retail Trade (Thousands of Persons) & 5 &  &  &  & X \\ 
  54 & USWTRADE & All Employees:  Wholesale Trade (Thousands of Persons) & 5 &  &  &  & X \\ 
  55 & CES9091000001 & All Employees:  Government:  Federal (Thousands of Persons) & 5 &  &  &  & X \\ 
  56 & CES9092000001 & All Employees:  Government:  State Government (Thousands of Persons) & 5 &  &  &  & X \\ 
  57 & CES9093000001 & All Employees:  Government:  Local Government (Thousands of Persons) & 5 &  &  &  & X \\ 
  58 & CE16OV & Civilian Employment (Thousands of Persons) & 5 & X & X & X & X \\ 
  59 & CIVPART & Civilian Labor Force Participation Rate (Percent) & 2 &  &  &  & X \\ 
  60 & UNRATE & Civilian Unemployment Rate (Percent) & 2 & X & X & X & X \\ 
  61 & UNRATESTx & Unemployment Rate less than 27 weeks (Percent) & 2 &  &  &  & X \\ 
  62 & UNRATELTx & Unemployment Rate for more than 27 weeks (Percent) & 2 &  &  & X & X \\ 
  63 & LNS14000012 & Unemployment Rate - 16 to 19 years (Percent) & 2 &  &  &  & X \\ 
  64 & LNS14000025 & Unemployment Rate - 20 years and over, Men (Percent) & 2 &  &  &  & X \\ 
  65 & LNS14000026 & Unemployment Rate - 20 years and over, Women (Percent) & 2 &  &  &  & X \\ 
  66 & UEMPLT5 & Number of Civilians Unemployed - Less Than 5 Weeks (Thousands of Persons) & 5 &  &  &  & X \\ 
  67 & UEMP5TO14 & Number of Civilians Unemployed for 5 to 14 Weeks (Thousands of Persons) & 5 &  &  &  & X \\ 
  68 & UEMP15T26 & Number of Civilians Unemployed for 15 to 26 Weeks (Thousands of Persons) & 5 &  &  &  & X \\ 
  69 & UEMP27OV & Number of Civilians Unemployed for 27 Weeks and Over (Thousands of Persons) & 5 &  &  &  & X \\ 
  70 & AWHMAN & Average Weekly Hours of Production and Nonsupervisory Employees:  Manufacturing (Hours) & 1 &  &  & X & X \\ 
  71 & AWOTMAN & Average Weekly Overtime Hours of Production and Nonsupervisory Employees: Manufacturing (Hours) & 2 &  &  & X & X \\ 
  72 & HWIx & Help-Wanted Index & 1 &  &  &  & X \\ 
  73 & CES0600000007 & Average Weekly Hours of Production and Nonsupervisory Employees:  Goods-Producing & 2 & X & X & X & X \\ 
  74 & CLAIMSx & Initial Claims & 5 &  &  & X & X \\ 
  75 & HOUST & Housing Starts: Total: New Privately Owned Housing Units Started & 5 &  & X & X & X \\ 
  76 & HOUST5F & Privately Owned Housing Starts: 5-Unit Structures or More & 5 &  &  &  & X \\ 
  77 & HOUSTMW & Housing Starts in Midwest Census Region (Thousands of Units) & 5 &  &  &  & X \\ 
  78 & HOUSTNE & Housing Starts in Northeast Census Region (Thousands of Units) & 5 &  &  &  & X \\ 
  79 & HOUSTS & Housing Starts in South Census Region (Thousands of Units) & 5 &  &  &  & X \\ 
  80 & HOUSTW & Housing Starts in West Census Region (Thousands of Units) & 5 &  &  &  & X \\ 
  81 & RSAFSx & Real Retail and Food Services Sales (Millions of Chained 2012 Dollars) & 5 &  &  & X & X \\ 
  82 & AMDMNOx & Real Manufacturers New Orders:  Durable Goods (Millions of 2012 Dollars) & 5 &  &  &  & X \\ 
  83 & AMDMUOx & Real Value of Manufacturers Unfilled Orders for Durable Goods Industries & 5 &  &  &  & X \\ 
  84 & BUSINVx & Total Business Inventories (Millions of Dollars) & 5 &  &  &  & X \\ 
  85 & ISRATIOx & Total Business:  Inventories to Sales Ratio & 2 &  &  &  & X \\ 
  86 & PCECTPI & Personal Consumption Expenditures: Chain-type Price Index  & 6 & X & X & X & X \\ 
  87 & PCEPILFE & Personal Consumption Expenditures Excluding Food and Energy & 6 & X & X & X & X \\ 
  88 & GDPCTPI & Gross Domestic Product: Chain-type Price Index & 6 & X & X & X & X \\ 
  89 & GPDICTPI & Gross Private Domestic Investment: Chain-type Price Index  & 6 &  &  & X & X \\ 
  90 & IPDBS & Business Sector:  Implicit Price Deflator (Index 2012=100) & 6 &  &  & X & X \\ 
   \hline
\end{tabular}
}
\end{table}
\begin{table}
\scalebox{0.52}{
\begin{tabular}{lllccccc} 
  \hline
  ID & FRED Code & Description & Transformation Codes & S & M & L & XL \\ 
  \hline
  91 & DGDSRG3Q086SBEA & Personal consumption expenditures:  Goods  & 6 &  &  & X & X \\ 
  92 & DDURRG3Q086SBEA & Personal consumption expenditures:  Durable goods  & 6 &  &  & X & X \\ 
  93 & DSERRG3Q086SBEA & Personal consumption expenditures:  Services  & 6 &  &  & X & X \\ 
  94 & DNDGRG3Q086SBEA & Personal consumption expenditures:  Nondurable goods & 6 &  &  & X & X \\ 
  95 & DHCERG3Q086SBEA & Personal consumption expenditures:  Services:  Household consumption expenditures & 6 &  &  &  & X \\ 
  96 & DMOTRG3Q086SBEA & Personal consumption expenditures:  Durable goods:  Motor vehicles and parts & 6 &  &  &  & X \\ 
  97 & DFDHRG3Q086SBEA & Personal consumption expenditures:  Durable goods:  Furnishings and durable household equipment & 6 &  &  &  & X \\ 
  98 & DREQRG3Q086SBEA & Personal consumption expenditures:  Durable goods:  Recreational goods and vehicles & 6 &  &  &  & X \\ 
  99 & DODGRG3Q086SBEA & Personal consumption expenditures:  Durable goods:  Other durable goods & 6 &  &  &  & X \\ 
  100 & DFXARG3Q086SBEA & Personal consumption expenditures:  Nondurable goods:  Food and beverages purchased for off-premises consumption & 6 &  &  &  & X \\ 
  101 & DCLORG3Q086SBEA & Personal consumption expenditures:  Nondurable goods:  Clothing and footwear & 6 &  &  &  & X \\ 
  102 & DGOERG3Q086SBEA & Personal consumption expenditures:  Nondurable goods:  Gasoline and other energy goods & 6 &  &  &  & X \\ 
  103 & DONGRG3Q086SBEA & Personal consumption expenditures:  Nondurable goods:  Other nondurable goods & 6 &  &  &  & X \\ 
  104 & DHUTRG3Q086SBEA & Personal consumption expenditures:  Services:  Housing and utilities & 6 &  &  &  & X \\ 
  105 & DHLCRG3Q086SBEA & Personal consumption expenditures:  Services:  Health care & 6 &  &  &  & X \\ 
  106 & DTRSRG3Q086SBEA & Personal consumption expenditures:  Transportation services & 6 &  &  &  & X \\ 
  107 & DRCARG3Q086SBEA & Personal consumption expenditures: Recreation services & 6 &  &  &  & X \\ 
  108 & DFSARG3Q086SBEA & Personal consumption expenditures:  Services:  Food services and accomodations & 6 &  &  &  & X \\
  109 & DIFSRG3Q086SBEA & Personal consumption expenditures:  Financial services and insurance & 6 &  &  &  & X \\ 
  110 & DOTSRG3Q086SBEA & Personal consumption expenditures:  Other services  & 6 &  &  &  & X \\ 
  111 & CPIAUCSL & Consumer Price Index for All Urban Consumers:  All Items & 6 & X & X & X & X \\ 
  112 & CPILFESL & Consumer Price Index for All Urban Consumers:  All Items Less Food \& Energy & 6 & X & X & X & X \\ 
  113 & WPSFD49207 & Producer Price Index by Commodity for Finished Goods  & 6 &  &  & X & X \\ 
  114 & PPIACO & Producer Price Index for All Commodities  & 6 &  &  & X & X \\ 
  115 & WPSFD49502 & Producer Price Index by Commodity for Finished Consumer Goods  & 6 &  &  &  & X \\ 
  116 & WPSFD4111 & Producer Price Index by Commodity for Finished Consumer Foods & 6 &  &  &  & X \\ 
  117 & PPIIDC & Producer Price Index by Commodity Industrial Commodities  & 6 &  &  &  & X \\ 
  118 & WPSID61 & Producer Price Index by Commodity Intermediate Materials:  Supplies \& Components & 6 &  &  &  & X \\ 
  119 & WPU0561 & Producer Price Index by Commodity for Fuels and Related Products and Power & 5 &  &  & X & X \\ 
  120 & OILPRICEx & Real Crude Oil Prices:  West Texas Intermediate (WTI) - Cushing, Oklahoma & 5 &  &  & X & X \\ 
  121 & WPSID62 & Producer Price Index:  Crude Materials for Further Processing  & 6 &  &  &  & X \\ 
  122 & PPICMM & Producer Price Index:  Commodities:  Metals and metal products:  Primary nonferrous metals & 6 &  &  &  & X \\ 
  123 & CPIAPPSL & Consumer Price Index for All Urban Consumers:  Apparel & 6 &  &  & X & X \\ 
  124 & CPITRNSL & Consumer Price Index for All Urban Consumers:  Transportation & 6 &  &  & X & X \\ 
  125 & CPIMEDSL & Consumer Price Index for All Urban Consumers:  Medical Care & 6 &  &  & X & X \\ 
  126 & CUSR0000SAC & Consumer Price Index for All Urban Consumers:  Commodities & 6 &  &  & X & X \\ 
  127 & CES2000000008x & Real Average Hourly Earnings of Production and Nonsupervisory Employees: Construction & 5 &  &  & X & X \\ 
  128 & CES3000000008x & Real Average Hourly Earnings of Production and Nonsupervisory Employees: Manufacturing & 5 &  &  & X & X \\ 
  129 & COMPRNFB & Nonfarm Business Sector:  Real Compensation Per Hour (Index 2012=100) & 5 &  &  & X & X \\ 
  130 & CES0600000008 & Average Hourly Earnings of Production and Nonsupervisory Employees: & 6 & X & X & X & X \\ 
  131 & FEDFUNDS & Effective Federal Funds Rate (Percent) & 2 & X & X & X & X \\ 
  132 & TB3MS & 3-Month Treasury Bill: Secondary Market Rate (Percent) & 2 &  &  & X & X \\ 
  133 & TB6MS & 6-Month Treasury Bill: Secondary Market Rate (Percent) & 2 &  &  & X & X \\ 
  134 & GS1 & 1-Year Treasury Constant Maturity Rate (Percent) & 2 &  & X & X & X \\ 
  135 & GS10 & 10-Year Treasury Constant Maturity Rate (Percent) & 2 & X & X & X & X \\ 
  136 & AAA & Moody's Seasoned Aaa Corporate Bond Yield (Percent) & 2 &  &  &  & X \\ 
  137 & BAA & Moody's Seasoned Baa Corporate Bond Yield (Percent) & 2 &  &  & X & X \\ 
  138 & BAA10YM & Moody's Seasoned Baa Corporate Bond Yield Relative to Yield on 10-Year Treasury & 1 &  &  &  & X \\ 
  139 & TB6M3Mx & 6-Month Treasury Bill Minus 3-Month Treasury Bill, secondary market (Percent) & 1 &  &  & X & X \\ 
  140 & GS1TB3Mx & 1-Year Treasury Constant Maturity Minus 3-Month Treasury Bill, secondary market & 1 &  &  & X & X \\ 
  141 & GS10TB3Mx & 10-Year Treasury Constant Maturity Minus 3-Month Treasury Bill, secondary market & 1 &  &  & X & X \\ 
  142 & CPF3MTB3Mx & 3-Month Commercial Paper Minus 3-Month Treasury Bill, secondary market & 1 &  &  & X & X \\ 
  143 & GS5 & 5-Year Treasury Constant Maturity Rate & 2 &  &  & X & X \\ 
  144 & TB3SMFFM & 3-Month Treasury Constant Maturity Minus Federal Funds Rate & 1 &  &  & X & X \\ 
  145 & T5YFFM & 5-Year Treasury Constant Maturity Minus Federal Funds Rate & 1 &  &  & X & X \\ 
  146 & AAAFFM & Moody's Seasoned Aaa Corporate Bond Minus Federal Funds Rate & 1 &  &  & X & X \\ 
  147 & M1REAL &  Real M1 Money Stock & 5 &  &  & X & X \\ 
  148 & M2REAL & Real M2 Money Stock & 5 &  & X & X & X \\ 
  149 & BUSLOANSx & Real Commercial and Industrial Loans, All Commercial Banks & 5 &  &  & X & X \\ 
  150 & CONSUMERx & Real Consumer Loans at All Commercial Banks  & 5 &  &  & X & X \\ 
  151 & NONREVSLx & Total Real Nonrevolving Credit Owned and Securitized, Outstanding & 5 &  &  & X & X \\ 
  152 & REALLNx & Real Real Estate Loans, All Commercial Banks & 5 &  &  & X & X \\ 
  153 & TOTALSLx & Total Consumer Credit Outstanding & 5 &  &  & X & X \\ 
  154 & TOTRESNS & Total Reserves of Depository Institutions  & 6 &  & X & X & X \\ 
  155 & NONBORRES & Reserves Of Depository Institutions, Nonborrowed & 7 &  & X & X & X \\ 
  156 & DTCOLNVHFNM & Consumer Motor Vehicle Loans Outstanding Owned by Finance Companies & 6 &  &  &  & X \\ 
  157 & DTCTHFNM & Total Consumer Loans and Leases Outstanding Owned and Securitized by Finance Companies  & 6 &  &  &  & X \\ 
  158 & INVEST & Securities in Bank Credit at All Commercial Banks  & 6 &  &  &  & X \\ 
  159 & TABSHNOx & Real Total Assets of Households and Nonprofit Organizations & 5 &  &  &  & X \\ 
  160 & EXSZUSx & Switzerland / U.S. Foreign Exchange Rate & 5 &  &  & X & X \\ 
  161 & EXJPUSx & Japan / U.S. Foreign Exchange Rate & 5 &  &  & X & X \\ 
  162 & EXUSUKx & U.S. / U.K. Foreign Exchange Rate & 5 &  &  & X & X \\ 
  163 & EXCAUSx & Canada / U.S. Foreign Exchange Rate & 5 &  &  & X & X \\ 
  164 & S.P.500 & S\&P's Common Stock Price Index:  Composite & 5 &  & X & X & X \\ 
  165 & S.P..indust & S\&P'sCommon Stock Price Index:  Industrials & 5 &  &  &  & X \\ 
  166 & S.P.div.yield & S\&P's Composite Common Stock:  Dividend Yield & 2 &  &  &  & X \\ 
   \hline
\end{tabular}
}
   \smallskip
\begin{minipage}{\linewidth}\small
\tiny \textbf{Notes}: This table provides an overview of the dataset employed. The transformation codes are applied to each time series $\bm Y_j$ and described in \cite{mccracken2020fred}: (1) no transformation; (2) $\Delta y_{jt}$; (3) $\Delta^2 y_{jt}$; (4) $\log (y_{jt})$; (5) $\Delta \log (y_{jt})$; (6) $\Delta^2 \log (y_{jt})$; (7) $\Delta (y_{jt}/y_{jt-1} - 1)$. 'X' marks the inclusion of a variable into one of the datasets.
\end{minipage}
\end{table}

\end{appendices}

\end{document}